\documentclass[prb,twocolumn,longbibliography,superscriptaddress,preprintnumbers]{revtex4-2}
\usepackage[colorlinks,bookmarks=true,citecolor=blue,linkcolor=blue,urlcolor=blue]{hyperref}
\usepackage{dcolumn,graphicx,amsfonts,amsthm,bm,color,appendix,float}

\usepackage{braket}
\usepackage[normalem]{ulem}
\usepackage[version=4]{mhchem}
\usepackage{siunitx}
\usepackage{amsmath}
\usepackage{amssymb}
\usepackage{CJKutf8}
\usepackage{mathtools}
\usepackage{enumitem}
\usepackage{booktabs}
\usepackage{bbold}

\usepackage[dvipsnames]{xcolor}
\definecolor{pal0}{rgb}{0.8941, 0.102 , 0.1098}
\definecolor{pal1}{rgb}{0.2157, 0.4941, 0.7216}
\definecolor{pal2}{rgb}{0.302 , 0.6863, 0.2902}
\definecolor{pal3}{rgb}{0.5961, 0.3059, 0.6392}
\definecolor{pal4}{rgb}{1.    , 0.498 , 0.    }

\newcommand{\n}[1]{\left| #1 \right|}
\newcommand{\st}[1]{\left\{#1\right\}}
\newcommand{\vv}[1]{\boldsymbol{#1}}

\renewcommand{\k}{\vv{k}}
\newcommand{\q}{\vv{q}}
\renewcommand{\r}{\vv{r}}
\newcommand{\g}{\vv{g}}

\begin{document}

\title{
Anomalous Hall Crystals in Rhombohedral Multilayer Graphene II: General Mechanism and a Minimal Model
}

\author{Tomohiro Soejima (\begin{CJK*}{UTF8}{bsmi}副島智大\end{CJK*})}
\thanks{These authors contributed equally.}
\affiliation{Department of Physics, Harvard University, Cambridge, MA 02138, USA}

\author{Junkai Dong (\begin{CJK*}{UTF8}{bsmi}董焌\end{CJK*}\begin{CJK*}{UTF8}{gbsn}锴\end{CJK*})}
\thanks{These authors contributed equally.}
\affiliation{Department of Physics, Harvard University, Cambridge, MA 02138, USA}

\author{Taige Wang}
\affiliation{Department of Physics, University of California, Berkeley, CA 94720, USA}
\affiliation{Material Science Division, Lawrence Berkeley National Laboratory, Berkeley, CA 94720, USA}
\affiliation{Kavli Institute for Theoretical Physics, University of California, Santa Barbara, California 93106, USA}

\author{Tianle Wang}
\affiliation{Department of Physics, University of California, Berkeley, CA 94720, USA}
\affiliation{Material Science Division, Lawrence Berkeley National Laboratory, Berkeley, CA 94720, USA}

\author{Michael P. Zaletel}
\affiliation{Department of Physics, University of California, Berkeley, CA 94720, USA}
\affiliation{Material Science Division, Lawrence Berkeley National Laboratory, Berkeley, CA 94720, USA}

\author{Ashvin Vishwanath}
\affiliation{Department of Physics, Harvard University, Cambridge, MA 02138, USA}

\author{Daniel E. Parker}
\affiliation{Department of Physics, University of California, Berkeley, CA 94720, USA}
\affiliation{Department of Physics, University of California at San Diego, La Jolla, California 92093, USA}

\begin{abstract}
We propose a minimal {``three-patch model"} for the anomalous Hall crystal (AHC), a topological electronic state that spontaneously breaks both time-reversal symmetry and continuous translation symmetry. The proposal for this state is inspired by the recently observed integer and fractional quantum Hall states in rhombohedral multilayer graphene at zero magnetic field. There, interaction effects appear to amplify the effects of a weak moir\'e potential, leading to the formation of stable, isolated Chern bands. It has been further shown that Chern bands are stabilized in mean field calculations even without a moir\'e potential, enabling a realization of the AHC state. Our model is built upon the dissection of the Brillouin zone into patches centered around high symmetry points. Within this model, the wavefunctions at high symmetry points fully determine the topology and energetics of the state. We extract two quantum geometrical phases of the non-interacting wavefunctions that control the stability of the topologically nontrivial AHC state. The model predicts that the AHC state wins over the topological trivial Wigner crystal in a wide range of parameters, and agrees very well with the results of full self-consistent Hartree-Fock calculations of the rhombohedral multilayer graphene Hamiltonian.
\end{abstract}

\maketitle


\section{Introduction}

Interacting electrons in the absence of external periodic potential can spontaneously break continuous translation symmetry, a phenomenon first predicted by Wigner in the context of the jellium model~\cite{wigner_interaction_1934}. Solid state systems, where translation symmetry is explicitly broken by the atomic potential, seem incompatible with realizing such spontaneous translation breaking. However, continuous translation symmetry can be (approximately) restored at very low electron densities: experiments both with strong magnetic field~\cite{lozovik_crystallization_1975,andrei_observation_1988,santos_observation_1992,yang_experimental_2021,tsui_direct_2023} and without~\cite{grimes_evidence_1979,yoon_wigner_1999,hossain_observation_2020,smolenski_signatures_2021,zhou_bilayer_2021,falson_competing_2022,sung_observation_2023,xiang_quantum_2024} have reported Wigner crystals. The profusion of two-dimensional materials with controllable electron densities
therefore presents an excellent opportunity to explore the physics of electron crystallization, yet little has been realized, or even predicted, beyond Wigner crystals.

While Wigner crystals are essentially classical crystals of localized electrons, it is in principle possible to generate a topologically nontrivial phase under spontaneous translation symmetry breaking. This was first borne out in \textit{Hall crystals}, a topologically nontrivial state formed by electrons in a Landau level spontaneously breaking translation symmetry
~\cite{kivelson_cooperative_1986,halperin_compatibility_1986,kivelson_cooperative_1987,tesanovic_hall_1989}.
Surprisingly, recent experiments on rhombohedral multilayer graphene (RMG)~\cite{lu_fractional_2024} suggested an avenue for realizing such a topological Wigner crystal at zero magnetic field.
There, pentalayer RMG, with a moir\'e potential induced by an aligned hBN crystal on one side, shows both Chern insulating (CI) and fractional Chern insulating (FCI) states at zero magnetic field, also called fractional quantum anomalous Hall (FQAH) states~\cite{regnault_fractional_2011,tang_high-temperature_2011,sun_nearly_2011,neupert_fractional_2011,bergholtz_topological_2013,liu_recent_2024}.
These states appear under a sizable perpendicular displacement field that polarizes the electrons to one side --- and combined experimental~\cite{lu_fractional_2024,chen_tunable_2020} and theoretical evidence~\cite{dong_theory_2023,zhou_fractional_2023,dong_anomalous_2023,guo_theory_2023,kwan_moire_2023} indicate it is the side \textit{away} from the hBN moir\'e, strongly suppressing the moir\'e potential. This creates a puzzle: how can such a weak moir\'e potential stabilize a Chern insulator?

Remarkably, self-consistent Hartree-Fock (SCHF) calculations in models of RMG revealed that the Chern insulator is stable even in the absence of moir\'e potential~\cite{zhou_fractional_2023,dong_anomalous_2023}. In this limit the state is interpreted as spontaneously breaking translation symmetry. To emphasize that it appears at \textit{zero} magnetic field, the state was named an \textit{anomalous} Hall crystal.
In mean-field treatments of models without the moir\'{e} potential, the state is surprisingly stable to various perturbations, such as changes in the microscopic hopping parameters. 
The role of the moir\'e potential in experiment is currently under debate~\cite{kwan_moire_2023}. However, the ubiquity of the AHC phase in theoretical models without a moir\'e potential is in strong contrast to the case of strong moir\'{e} limit.

The appearance of AHC ground states in these models poses a fundamental conceptual question: when and why do these models give rise to a topological state? Due to its nontrivial topology, the AHC state does not admit a simple classical explanation for its origin \'{a} la Wigner crystals. In this paper, we shed light on a simple yet fully quantum-mechanical mechanism underlying the origin of the AHC ground states.

Beyond the fundamental interest in its stability, the AHC provides a mechanism to generate nontrivial band topology different from previous examples such as nonzero magnetic field, band inversion~\cite{bernevig_quantum_2006,fu_topological_2007,liu_quantum_2008,zhang_topological_2009} and background skyrmionic textures~\cite{ye_berry_1999,ohgushi_spin_2000,hamamoto_quantized_2015,van_hoogdalem_magnetic_2013,wu_topological_2019,paul_topological_2021,divic_magnetic_2022}. Here, it is the spontaneously broken translation symmetry that predominantly generates the requisite Berry curvature.
The source of this Berry curvature is the interference of plane waves: after translation breaking, the state at each crystalline momentum is a superposition of several plane waves whose phase winding generates Berry curvature~\cite{bohm2003geometric}.
In fact, the Chern number is highly constrained by the interference pattern between a small subset of plane waves near the high-symmetry points.

To study the stability and origin of topology,
we construct a minimal phenomenological model, the ``three-patch model,'' wherein we keep a minimal number of plane waves to represent translation-broken states with different Chern numbers. This amounts to keeping only three electrons and seven plane wave states.

Despite its simplicity, the three-patch model both illustrates when the AHC state is energetically favorable, and matches realistic phase diagrams closely. Particularly, we show that the stability of the AHC phase depends on the scattering phases between plane waves, which are in turn dictated by the quantum geometric phases of the plane wave states. Furthermore, the three-patch model reproduces the full Hartree-Fock prediction for the phase boundary between the AHC and a trivial crystal with striking accuracy, serving as \textit{a posteriori} confirmation of its generality and applicability. The three-patch model thus gives a tractable way to understand the anomalous Hall crystal mechanism of generating quantized anomalous Hall states.

The remainder of this work is organized as follows. In Section \ref{sec:three-patch-model} we introduce the ``three-patch model", an analytical-solvable model with a stable anomalous Hall crystal phase. Based on this simple picture, we make a number of specific predictions about microscopic models of RMG.
By studying the model with RMG spinors, we predict that both the Hartree and Fock interactions cooperate to stabilize the state in a large part of the parameter space. We further predict a quantum phase transition between the $C=0$ state and $C=1$ state, driven by a change in quantum geometry.

We then provide an intuitive picture to understand the charge densities of the three-patch states. This also allows us to graphically examine the energetics of the states, which reveals that the Hartree interaction always favors the AHC state.
In Section \ref{sec:RMG}, we perform SCHF calculation of a microscopic Hamiltonian of RMG. We confirm that ground states found in SCHF match the three-patch characteristics, and verify that the SCHF phase diagram is closely mirrored by the prediction by the three-patch model.
We conclude with some discussion about future directions in Section \ref{sec:discussion}.

\section{A Three-patch model of an anomalous Hall crystal}
\label{sec:three-patch-model}

We discuss a mechanism for generating Chern bands in models of interacting electrons where both crystallinity, i.e. spontaneous breaking of the continuous translation, {\em and} band topology are spontaneously generated. Some of the ingredients, which are made more precise below, are (i) a relatively flat energy dispersion in the momentum range of interest, and a large dispersion outside it, (ii) quantum geometry encoded in a multicomponent wavefunction that varies over the Brillouin zone, which seeds the Chern character of the final state and (iii) $C_3$ symmetry, retained partly for analytical convenience. This general mechanism is encapsulated in a minimal ``three-patch model". Solving the minimal model gives a simple criterion based on quantum geometry for obtaining the AHC as the ground state, which is satisfied in a wide range of parameters.

\subsection{Construction of the Three-Patch Model}
\label{subsec:three_patch_construction}
We now construct the three-patch model, highlighting the hypotheses it requires.

Consider an electronic band of a multi-orbital continuum quantum system with continuous translation symmetry, $C_3$ symmetry, and broken time-reversal symmetry (for example, the $K$ valley of rhombohedral multilayer graphene). The Hilbert space is spanned by plane wave states $\ket{\vv{q}, a}$, where $\vv{q}$ is the wavevector and $a$ denotes the component (i.e. spin, layer, sublattice, etc.). We choose a basis for components $\ket{a}$ that are irreps of $C_3$, so that the action of $C_3$ on the basis states is
\begin{equation}
    \hat{C}_3 \ket{\vv{q}, a} = \exp\left[\frac{2\pi i\ell_a}{3}\right] \ket{C_3 \vv{q}, a},\quad \ell_a \in \mathbb{Z}/3 \mathbb{Z}.
\end{equation}
Eigenstates $\hat{c}_{\q}^\dagger \ket{0} = \ket{\phi_{\q}}$ in the band of interest are indexed by unrestricted momentum $\vv{q}$ such that
\begin{equation}\label{eq:singleparticlestates}
    \hat{h} \ket{\phi_{\q}}  = E_{\q} \ket{\phi_{\q}}, \quad 
    \hat{T}_{\r} \ket{\phi_{\q}} = e^{i\q \cdot \r} \ket{\phi_{\q}},
\end{equation}
where $\hat{h}$ is the single-particle Hamiltonian of the system, and $\hat{T}_{\r'} \phi(\r) = \phi(\r+\r')$. These wavefunctions therefore take the unnormalized plane wave form
\begin{equation}
    \phi_{\q}^a(\r) = \braket{\r, a| \phi_{\q}} =  \chi^{a}({\q}) e^{i \q\cdot\vv{r}}.
\end{equation}
We also define $\chi(\vv q) = \sum_{a} \chi^a({\q}) \ket{a}$ to be the orbital-space spinor.

We add density-density interactions in the usual way:
\begin{equation}
\label{eq:hamiltonian}
\begin{aligned}
    \hat{H}_0 &= \hat{h} + \hat{H}_{\textrm{int}}\\
    & = \hat{h}+\frac{1}{2} \int d^2\r\; d^2\r'\; V(\r -\r') : \hat{\rho}(\r) \hat{\rho}(\r'):,
\end{aligned}
\end{equation}
where $\hat{\rho}(\vv{r})$ is the sum of real space densities over all components. 

\begin{figure}
    \centering
    \includegraphics[width=\linewidth]{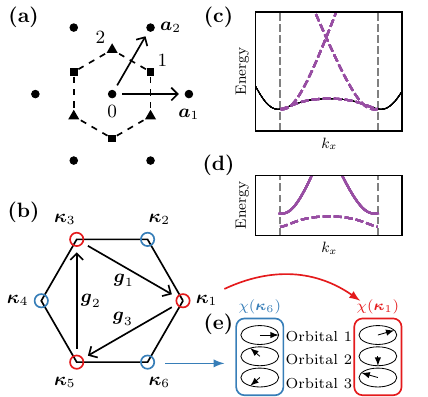}
    \caption{Setup of the three-patch model. (a) Real space geometry of the superlattice. The three inequivalent Wyckoff positions are labeled by circles, squares, and triangles.
    (b) Mini Brillouin zone (mBZ) and associated high symmetry points. The points $\vv\kappa_1$, $\vv\kappa_3$, $\vv\kappa_5$ span the $\kappa$ point, while $\vv\kappa_2$, $\vv\kappa_4$, and $\vv\kappa_6$ span the $\kappa'$ point.
    (c) Schematic folded band structure of the model in the mBZ. The $\kappa$ and $\kappa'$ points have three-fold degeneracies. (d) Band gaps are opened up along mBZ boundaries by interactions, which crucially depend on (e) the nontrivial quantum geometry of multi-component wavefunctions.}
    \label{fig:3point-model-cartoon}
\end{figure}

We restrict our attention to a fixed electron density $n$, where we will study the energetic competition between crystalline insulating ground states. In anticipation of spontaneous translation symmetry breaking, we consider a hexagonal ``superlattice" generated by $\vv{a}_1,\vv{a}_2$, shown in  Fig.~\ref{fig:3point-model-cartoon}(a). The superlattice size is chosen such that there is one electron per unit cell i.e. filling $\nu=1$.  While this is \textit{not} due to an explicitly translation-breaking superlattice potential (e.g. moir\'e potential), we adopt notation from that context. The band structure is folded (but not yet electronically hybridized) onto a hexagonal mini Brillouin zone (mBZ), Fig.~\ref{fig:3point-model-cartoon}(b), with reciprocal lattice vectors $\vv{g}_1, \vv{g}_2$ and three high-symmetry points ${\gamma}, {\kappa},$ and ${\kappa}'$. Fig~\ref{fig:3point-model-cartoon}(b) shows $\vv{\kappa}_1, \vv{\kappa}_3,$ and $\vv{\kappa}_5$ get folded to ${\kappa}$, while $\vv{\kappa}_2, \vv{\kappa}_4$ and $\vv{\kappa}_6$ are folded to ${\kappa}'$.
We choose $\vv\gamma$ at a $C_3$ high symmetry point before translation breaking. The folded single-particle band structure is three-fold degenerate on the edge of the bottom band at ${\kappa}$ and ${\kappa}'$, which we call mini-valleys. The $C_3$ symmetry enforce degeneracy of the kinetic energy $E_{\vv{\kappa}_j} = E_{\vv{\kappa}_{j+2}}$, so the folded band structure is three-fold degenerate.

To make analytic progress, we impose three physical conditions on the ground states:
\begin{enumerate}
\item $C_3$-preserving Crystallization: spontaneous symmetry breaking of continuous translations while preserving $C_3$. This gives a hexagonal ``superlattice" with three $C_3$ invariant momenta ${\gamma}, {\kappa}$, and ${\kappa}'$.
\item Weak coupling mean field: the ground state is a Slater determinant state that only hybridizes energetically proximate bands.
\item Three-patch assumption: the single-particle orbitals that make up the ground state wavefunction is essentially homogeneous in large patches around the three $C_3$ invariant momenta.
This is in spirit similar to previous ``patch models" or ``hotspot models" in the superconductivity community~\cite{Hertz,Millis,Chubukov_patch,Chubukov_patch2,Chubukov_patch3,Max_patch,Subir_patch,Schulz_VHS_1987,Dzialoshinskii_VHS_1987,Poilblanc_1987_VHS,Manfred_VHS_1998,HUR20091452,Chubukov2012,Liang2018,TaigeRG}.
\end{enumerate}
We will argue in App.~\ref{sec:justifications} that these three conditions can be physically motivated if the band dispersion has a flat bottom and a dispersive edge. Moreover, we verify these three conditions are consistent with SCHF ground states of microscopic model of RMG in Sec.~\ref{sec:RMG}.

Under these assumptions, the ground states of $\hat{H}_0$ can be physically modeled by $C_3$ symmetric Slater determinants with one electron at each of the high symmetry points. This procedure is detailed in App.~\ref{sec:justifications}. Briefly, this follows by taking a single representative $\vv k$-point for each high-symmetry patch, which we conveniently take to be the high-symmetry points. Since the state is weakly coupled, only low-energy bands are allowed to hybridize, giving three states at ${\kappa}, {\kappa}'$ due to symmetry-enforced band crossings, but generically only one at ${\gamma}$, as shown in Fig.~\ref{fig:3point-model-cartoon}(c). The only possible single-particle wavefunctions are
\begin{align}
\label{eq:three_patch_wavefunctions}
    \psi_{\gamma,\ell_{\gamma}=0}(\vv{r}) &= \chi(\gamma) \\
    \nonumber 
    \psi_{\kappa,\ell_{\kappa}}(\vv r) &= \frac{1}{\sqrt{3}}\sum_{j=1,3,5} 
    \exp{\left[\frac{2\pi i \ell_{\kappa}}{3}\frac{j}{2}\right]}
    e^{i\vv \kappa_j\cdot \vv r}\chi(\vv \kappa_j),\\
    \nonumber
    \psi_{\kappa',\ell_{\kappa'}}(\vv r) &= \frac{1}{\sqrt{3}}\sum_{j=2,4,6} 
    \exp{\left[\frac{2\pi i \ell_{\kappa'}}{3}\frac{j}{2}\right]}
    e^{i\vv \kappa_j\cdot \vv r}\chi(\vv \kappa_j),
\end{align}
where we fix the gauge $\chi(\vv\kappa_{j+2}) = \hat{C}_3 \chi(\vv\kappa_{j})$, and the states are indexed by $C_3$ angular momenta $\ell_{\xi}$ defined by
\begin{equation}
    C_3 \ket{\psi_{\xi,\ell_{\xi}}} = \exp\left[\frac{2\pi i \ell_{\xi}}{3}\right] \ket{\psi_{\xi,\ell_{\xi}}}.
\end{equation}
In the above, we fixed the gauge such that $\ell_{\gamma} =0$.

Given the discrete choices at each momentum point, there are exactly $9$ Slater determinant states $\ket{\Psi_{\ell_{{\kappa}}, \ell_{{\kappa}'}}}$ at filling one electron per $k$-point, which are fully characterized by $\ell_{{\kappa}}$ and $ \ell_{{\kappa}'}$. In first quantization,
\begin{equation}
    \label{eq:three_patch_Slater_det}
    \Psi_{\ell_{{\kappa}}, \ell_{{\kappa}'}}(\vv{r}_1, \vv{r}_2, \vv{r}_3) = \mathcal{A}[\psi_{\gamma}(\vv{r}_1), \psi_{ \kappa,\ell_{\kappa}}(\vv{r}_2),  \psi_{\kappa',\ell_{\kappa'}}(\vv{r}_3)],
\end{equation}
where $\mathcal{A}$ is the antisymmetrization operator. Each $\ket{\Psi_{\ell_{{\kappa}}, \ell_{{\kappa}'}}}$ captures the character of a full 2D state whose Chern number is given by
\begin{equation}
    \label{eq:symmetry-indicators}
    C \equiv \ell_{{\kappa}} + \ell_{{\kappa}'} \mod 3.
\end{equation}
This follows from the
theory of symmetry indicators~\cite{fu_topological_2007,turner_quantized_2012,fang_bulk_2012,po_symmetry-based_2017,bradlyn_topological_2017,po_symmetry_2020,cano_band_2021}, 
which posits that the sum of angular momentum at $C_3$ symmetric momenta determines the Chern number mod $3$ i.e. $C\equiv \ell_{{\gamma}} + \ell_{{\kappa}} + \ell_{{\kappa}'}$. This can be thought of as a generalization of the famous Fu-Kane formula for inversion symmetric topological insulators~\cite{fu_topological_2007}, which was later generalized to Chern bands with discrete rotation symmetries~\cite{turner_quantized_2012, fang_bulk_2012}.  We choose the $C_3$ origin such that $\ell_{{\gamma}} = 0$ above, implying Eq.~\eqref{eq:symmetry-indicators}. We will refer to three possible values of $C$ as simply $C=1,0,-1$, although in principle the Chern number can be larger.

Due to $C_3$, these Slater determinant states have degenerate kinetic energy, so their energetic competition is determined solely through interactions:
\begin{equation}
    E(\ell_{{\kappa}}, \ell_{{\kappa'}}) = \braket{\Psi_{\ell_{{\kappa}}, \ell_{{\kappa}'}}|\hat{H}_\mathrm{int}|\Psi_{\ell_{{\kappa}}, \ell_{{\kappa}'}}}.
    \label{eq:three-patch-energy}
\end{equation}
Below we will evaluate the energies $E(\ell_{{\kappa}}, \ell_{{\kappa'}})$ analytically to determine when a $C=1$ state is stabilized.


\begin{figure*}
    \centering
    \includegraphics[width = \textwidth]{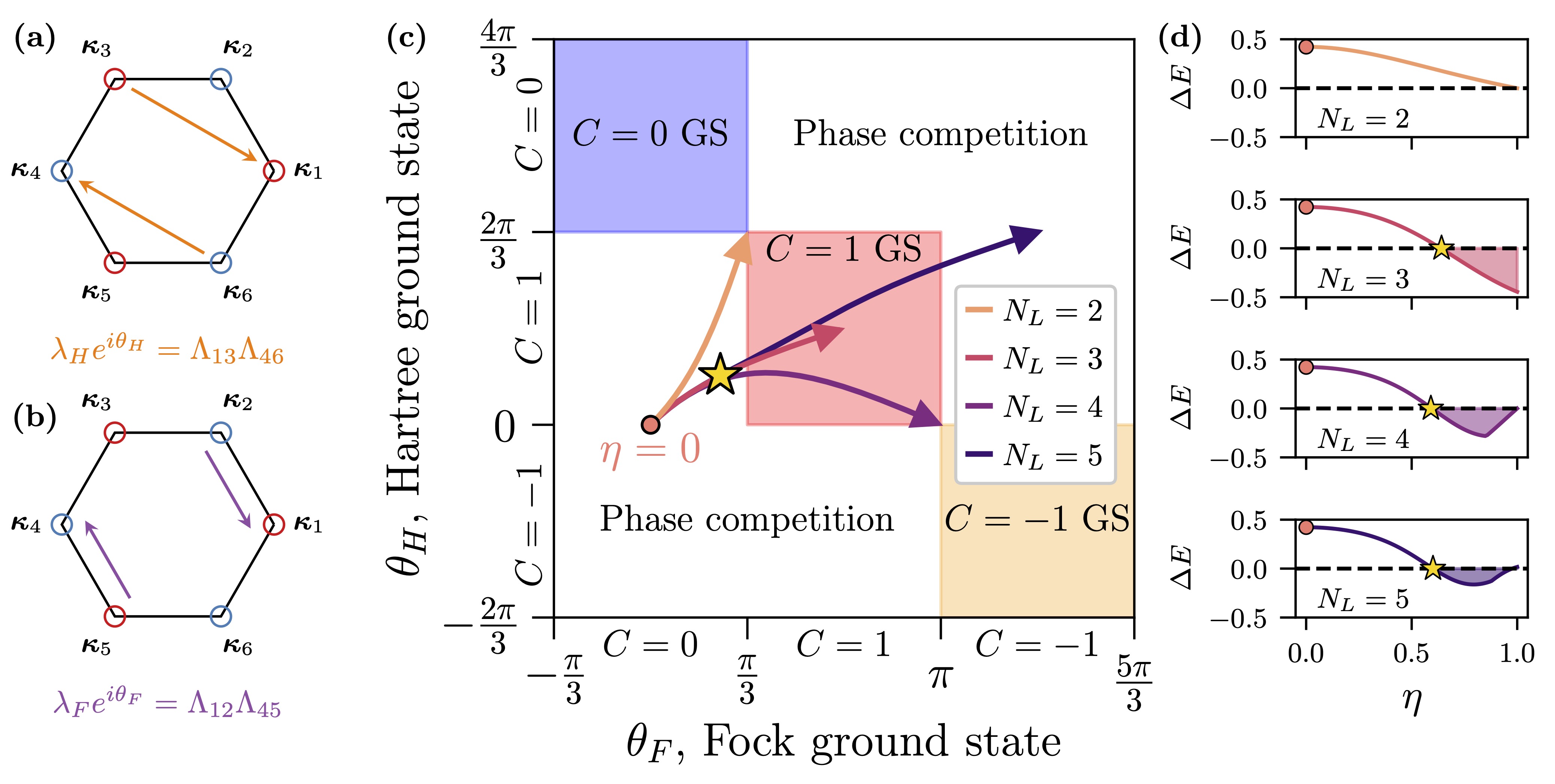}
    \caption{Predicted phase diagram by the three-patch model. (a) and (b) show the scattering processes contributing to the Hartree and Fock terms $E_{H}(C)$ and $E_F(C)$, and their associated form factors. (c) The angles of the form factors, $\theta_H$ and $\theta_F$, determine the Chern number favored by the Hartree and Fock terms respectively. Colored regions are where the Hartree and Fock ground states agree.  Curves correspond to the evolution of $\theta_H$ and $\theta_F$ as a function of $0\leq \eta \leq 1$, which tunes the RMG-inspired spinors $\chi^{(N_L)}$ for different number of components $N_L$,. The transition points where the $C=1$ state becomes the ground state in the three-patch model for $N_L=3,4,5$ are very close to each other, and are marked by the yellow star in the phase diagram. The orange dot marks the $\eta=0$ point. (d) The energy difference between the $C=1$ state and the lowest-energy $C\neq 1$ state in the three-patch model for different number of components $N_L$ as a function of $\eta$. Transition points to the $C=1$ states are marked by yellow stars.}
    \label{fig:phase_phase_diagram}
\end{figure*}

\subsection{Analytical solution of the three-patch model}
\label{subsec:analytical_solution}
We will now perform a full analytical evaluation of the three-patch energy in Eq.~\eqref{eq:three-patch-energy}. As it is the energy of a Slater determinant state, we can perform Wick contraction to separate Hartree and Fock contributions:

\begin{equation}\label{eq:wickterms}
\begin{aligned}
    E(\ell_{{\kappa}}, \ell_{{\kappa'}}) & = \frac{1}{2}\sum_{\substack{\xi,\xi' \in \\ \{\gamma, \kappa, \kappa'\}}} E_{H}(\xi,\xi') + E_{F}(\xi,\xi').
\end{aligned}
\end{equation}

Here, the Hartree and Fock (exchange) energies $E_H, E_F$ between single particle states are defined as
\begin{widetext}
    \begin{equation}
        E_H(\xi,\xi') = \int d^2\vv r d^2\vv r'
        \overline{\psi_{\xi}^a(\vv{r})}
        {\psi_{\xi}^a(\vv{r})}
        V(\vv r-\vv r')
        \overline{\psi_{\xi'}^b(\vv{r})}
        {\psi_{\xi'}^b(\vv{r})}, \quad
        E_F(\xi, \xi') = -\int d^2\vv r d^2\vv r' \overline{\psi_{\xi'}^{a}(\vv{r})} \psi_{\xi}^{a}(\vv{r}) V(\vv r-\vv r') \overline{\psi_{\xi}^{b}(\vv{r}')} \psi^{b}_{\xi'}(\vv{r}'),
    \end{equation}
\end{widetext}
where again $\psi^a_{\xi}(\vv r)$ is the component $a$ of the single-particle wavefunctions at high symmetry point $\xi$, and the repeated indiced are summed over.

Naively, the interaction energy Eq.~\eqref{eq:wickterms} comes with three pairs of terms. However, translation by $\tilde{\vv a}=(\vv a_1+\vv a_2)/3$ shifts angular momenta $\ell_{\kappa}\to \ell_{\kappa}+1$, and $\ell_{\kappa'}\to \ell_{\kappa'}-1$ as we show in App.~\ref{app:real_space_properties}. Thus, the energy can only depend on $\ell_{\kappa}+\ell_{\kappa'} \equiv C \mod 3$. Since the $(\xi, \xi')=(\gamma,\kappa)$ term does not depend on $\ell_{\kappa'}$, it cannot have dependence on angular momenta; the same goes for the $(\gamma,\kappa')$ term. The only term with dependence on $C$ is the $(\kappa,\kappa')$ term.

By explicit computation in Fourier space, 
we obtain the energy of the $(\kappa, \kappa')$ interaction term as a function of $\ell_{{\kappa}}, \ell_{{\kappa}'}$:
\begin{widetext}
    \begin{align}
        E_H(\kappa, \kappa') &= \frac{1}{9}\sum_{m, n \in \st{1, 3, 5}} \sum_{o, p \in \st{2, 4, 6}}
        e^{2\pi i\left[(n-m)\ell_{{\kappa}}+(p-o)\ell_{{\kappa}'}\right]/6}
        \Lambda_{m, n}
        \Lambda_{o, p}
        V_{\vv{\kappa}_n - \vv{\kappa}_m} \delta_{\vv{\kappa_n} - \vv{\kappa}_m, \vv{\kappa}_o -\vv{\kappa}_p},\\
        E_F(\kappa, \kappa') &= -\frac{1}{9} \sum_{m, n \in \st{1, 3, 5}} \sum_{o, p \in \st{2, 4, 6}}
        e^{2\pi i\left[(n-m)\ell_{{\kappa}}+(p-o)\ell_{{\kappa}'}\right]/6}
        \Lambda_{m, p}
        \Lambda_{o, n}
        V_{\vv{\kappa}_p - \vv{\kappa}_m} \delta_{\vv{\kappa_p} - \vv{\kappa}_m,  \vv{\kappa}_o - \vv{\kappa}_n},
    \end{align}
\end{widetext}
where we defined form factors $\Lambda_{m,n} = \chi^\dagger(\vv{\kappa}_m)\chi(\vv{\kappa}_n)$, and $V_{\vv{q}}$ is the Fourier transform of interaction $V(\vv{r})$ at momentum $\vv{q}$. The physical meaning of each term in this sum is clear: they correspond to different scattering processes with momentum transfer $\vv{\kappa}_m - \vv{\kappa}_n$. 

Only the scattering processes depicted in Fig.~\ref{fig:phase_phase_diagram}(a, b) and their $C_3$ rotated counterparts have $\ell_{{\kappa}}, \ell_{{\kappa}'}$ dependence. Collecting those together, and using the $C_3$ action $\Lambda_{n,m} = \Lambda_{n+2,m+2}$ to simplify the expression, we get
\begin{align}
    E(C) &= V_{\vv g_1}\epsilon_H(C) + V_{\vv \kappa_1}\epsilon_F(C) + E_0,\\
    \begin{split}\label{eq:Hartree_FF}
    \epsilon_H(C) &= \frac{2}{3}\mathrm{Re}\left[e^{2\pi iC/3}
    \Lambda_{1, 3}
    \Lambda_{4, 6}
    \right]\\
    &= \frac{2}{3} \lambda_H \cos\left(\theta_H+\frac{2\pi C}{3}\right)
    ,
    \end{split}\\
    \begin{split}\label{eq:Fock_FF}
    \epsilon_F(C) &= -\frac{2}{3}\mathrm{Re}\left[e^{-2\pi iC/3}
    \Lambda_{1, 2}
    \Lambda_{4, 5}
    \right]\\
    &=-\frac{2}{3}\lambda_F \cos\left(\theta_F - \frac{2\pi C}{3}\right).
    \end{split}
\end{align}
Here $E_0$ is a term that does not depend on the Chern number $C$. We parameterized the form factors in terms of two angles
\begin{equation}\label{eq:phase_HF}
    \Lambda_{1, 3}
\Lambda_{4, 6} = \lambda_H e^{i\theta_H},\quad
\Lambda_{1, 2}
\Lambda_{4, 5} =  \lambda_F e^{i\theta_F}.
\end{equation}
The ground state of $\epsilon_H(\epsilon_F)$ is determined by $\theta_H(\theta_F)$ alone. For certain values of $\theta_H$ and $\theta_F$, the ground states of the Hartree and the Fock terms agree, uniquely determining the ground state. In Fig.~\ref{fig:phase_phase_diagram}(c), we show a 2D phase diagram of the three-patch model as a function of $\theta_H$ and $\theta_F$. The ``frustration free" regions where the Hartree ground state and the Fock ground state agree are shaded with color.

We note these angles are connected to the Pancharatnam overlap
\begin{equation}\label{eq:pancharatnam}
    P = \prod_{i=1}^{6}\Lambda_{i-1,i} = \lambda_F^3 e^{3i\theta_F},
\end{equation}
where $\chi_0$ is defined to be $\chi_6$. The angle $\arg P$ records the accumulated geometric phase by a path along the mBZ boundaries. In the case where spinors vary slowly along the path, the phase $\arg P = 3\theta_F$ can be approximated by the total Berry curvature within the Brillouin zone. The \text{Fock term} starts favoring a $C=1$ state when $\theta_F > \pi/3$, which approximately translates to $\int_{\mathrm{mBZ}} \Omega(\vv k) \gtrsim \pi$,
i.e. half the Berry curvature necessary for the $C=1$ state. 
The \textit{Hartree} term, however, favors the $C=1$ state at arbitrarily small $\eta$, whose competition with the Fock term determines the ultimate ground state.

\subsection{Topological Transition from RMG spinors}
\label{subsec:three-patch-RMG}
\begin{figure*}
    \centering
    \includegraphics[width=\textwidth]{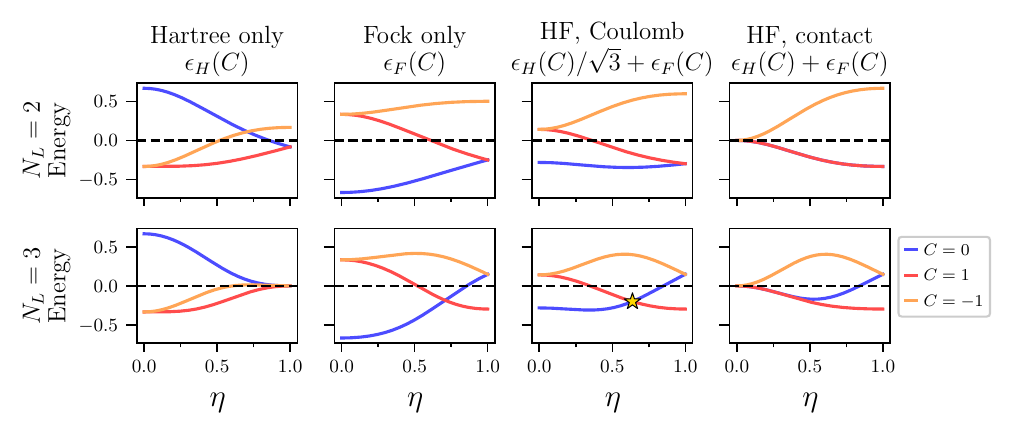}
    \caption{Hartree and Fock energies of different $C$ states. The rows correspond to different number of components. The first two columns correspond to the Hartree and Fock terms $E_H(C)$ and $E_F(C)$, plotted with $V_{\vv g} = V_{\vv\kappa} = 1$. ``HF, contact'' is the sum of Hartree and Fock energies with $V_{\vv{g}} = V_{\vv{\kappa}} = 1$, while ``HF, Coulomb'' is computed with $V_{\vv{\kappa}} = 1, V_{\vv{g}} = 1/\sqrt{3}$. The yellow star marks the phase transition point between $C=0$ and $C=1$ using Coulomb interaction.}
    \label{fig:three-patch_hartree-fock_comparison}
\end{figure*}

The phases $\theta_H$ and $\theta_F$ can depend on the detail of the spinor structure, which in turn determine whether the $C=1$ is stable. We now consider a concrete example of the spinor to examine the behavior of $\epsilon_H(C)$ and $\epsilon_F(C)$. We take the following spinor motivated by the physics of RMG systems:
\begin{equation}\label{eq:three-path-spinor}
    \chi^{(N_L)}(\vv\kappa_{n+1}) = \mathcal{N}(1,\eta e^{2\pi in/6}, \ldots, \eta^{N_L-1} e^{2\pi i(N_L-1)n/6}),
\end{equation}
where $\eta$ is a positive number less than 1, $\mathcal{N}=1/\sqrt{\sum_{a=0}^{N_L-1} \eta^{2a}}$ is the normalization factor. The components have $C_3$ action given by $\hat{C}_3=\mathrm{diag}(1,e^{2\pi i/3},\ldots,e^{2\pi i(N_L-1)/3})$. 

In Fig.~\ref{fig:phase_phase_diagram}(c), we show parametric curves
$(\theta_F(\eta), \theta_H(\eta))$ for $0\leq \eta \leq 1$ for various values of $N_L=2-5$. We see that at $N_L=2$, the curve never enters the region where both Hartree and Fock terms favor the $C=1$ ground state (labeled as ``$C=1$ GS'' in the figure), while for other number of layers, a significant portion of the curve is inside the ``$C=1$ GS'' region. We conclude that the AHC state forms at a large range of $\eta$ for $N_L=3-5$, regardless of the values of $V_{\vv{g}}$ and $V_{\vv{\kappa}}$.

To see this more quantitatively, we evaluate the Hartree and Fock energies explicitly. They can be organized in powers of $\eta$ (See App.~\ref{app:RMG_Hartree_Fock}):

\begin{align}
    \epsilon_H(C) &= \frac{2}{3} \sum_{a, b=0}^{N_L-1} \mathcal{N}^4 \eta^{a + b} \cos\left(\frac{2\pi (a+b+C)}{3}\right)\label{eq:RMG_Hartree} \\
    \epsilon_F(C) &= -\frac{2}{3} \sum_{a, b=0}^{N_L-1}\mathcal{N}^4 \eta^{a + b} \cos\left(\frac{2\pi (a+b-2C)}{6}\right).\label{eq:RMG_Fock}
\end{align}

We consider two types of interactions: short-range interaction, i.e. contact interaction $V(\vv r) \propto \delta(\vv r), V_{\vv{q}} = \textrm{const.}$ ($V_{\vv\kappa_1}=V_{\vv g_1}$), and long-range interaction, i.e. Coulomb interaction  $V(\vv r)\propto 1/|\vv r|, V_{\vv{q}} \propto 1/|\vv{q}|$ ($V_{\vv\kappa_1}=\sqrt{3}V_{\vv g_1}$).

Let us define $\Delta E$ to be the energy difference between the Coulomb energy of the $C=1$ state and the lowest energy state:

\begin{equation}
    \Delta E \coloneqq \frac{\epsilon_H(1)}{\sqrt{3}} +  \epsilon_F(1) - \min_{C=0, -1} \left[\frac{\epsilon_H(C)}{\sqrt{3}} +  \epsilon_F(C)\right].
\end{equation}
Values $\Delta E < 0$ signal that $C=1$ is the ground state with the Coulomb interaction. We plot this for $N_L = 2 - 5$  in Fig.~\ref{fig:phase_phase_diagram}(d). We see that for $N_L \geq 3$, $C=1$ becomes the ground state at $\eta \sim 0.6$, marked by a yellow star, while the $C=1$ never becomes the unique ground state for $N_L = 2$. The corresponding critical point is marked in
Fig.~\ref{fig:phase_phase_diagram}(c) as well, where the single yellow star covers all the critical points. As the Hartree term always favors the $C=1$ state, the transition point is slightly outside of the ``$C=1$ GS'' region. We conclude that the AHC state is favored for large enough $\eta$ for $N_L \geq 3$.

Having understood the case of Coulomb interaction for $N_L=2 - 5$, we now look at Hartree and Fock contributions separately. Since many of the qualitative features are similar for all $N_L\geq 3$, we focus on $N_L = 2, 3$. We plot the Hartree and Fock energies ($\epsilon_H(C), \epsilon_F(C)$), as well as contact and Coulomb energies ($\epsilon_H(C) + \epsilon_F(C), \epsilon_H(C)/\sqrt{3} + \epsilon_F(C)$) in Fig.~\ref{fig:three-patch_hartree-fock_comparison}.
We note the following features of these plots, which can also be analytically confirmed: 1) Hartree always favors the $C=1$ state, 2) Fock favors the $C=0$ for all values of $\eta$ for $N_L=2$, but favors $C=1$ above $\eta\approx 1/\sqrt{2}$ for $N_L\geq3$, 3) Coulomb interaction energy behaves qualitatively similarly to the Fock interaction energy, and 4) The contact interaction makes $C=0$ and $C=1$ exactly degenerate at $N_L=2$, which is a general feature of $N_L=2$ spinors (which is true in general for $N_L=2$, see App.~\ref{app:two_component_degenerate}).

We stress again that the energetics of the phase transition for this choice of spinors is mostly governed by the Fock exchange energy, manifesting in the similarity between the Fock energy curve and the Coulomb energy curve. While the Hartree energy always favors the $C=1$ state, it only shifts the phase transition point slightly.

\subsection{Three-patch states in real space}
\label{sec:graphical_intuition_and_real_space}
In this section, we inspect the charge density of the three-patch states with RMG spinors, discussed in Sec.~\ref{subsec:three-patch-RMG}. These charge densities are not only characteristic features of the different three-patch states, but also let us peek into their energetics in different regimes.

\begin{figure}
    \centering
        \includegraphics[width=\linewidth]{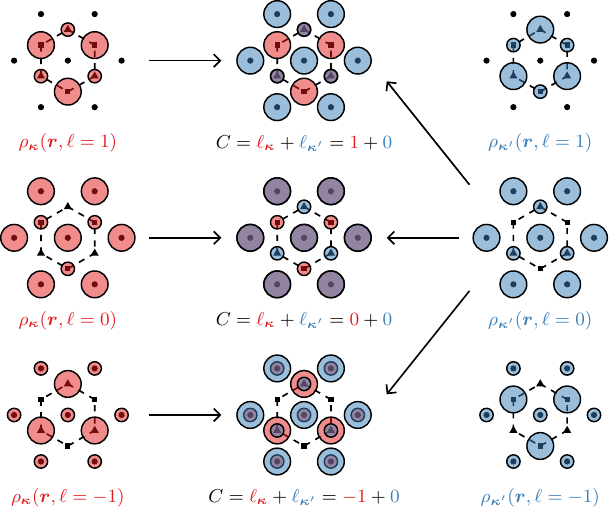}
    \caption{Charge densities of competing candidate states of the three patch model. (left): the charge densities at the $\kappa$ point. The size of the circles corresponds to different amplitudes of charge densities, which come from different components of the spinor. Different angular momentum states labeled by $\ell$ are related to each other by translation. (right): the charge densities in the ${\kappa}'$ point. (center): the total charge densities of states with different Chern numbers.}
    \label{fig:3point-model-hartree}
\end{figure}

Let us consider the single particle state at ${\kappa}$ with angular momentum $\ell_{\kappa}$.  The $C_3$ symmetry of the spinors $\hat{C_3} \chi(\vv{\kappa}_i) = \chi(\vv{\kappa}_{i+2})$, fixes the $i$ dependence of
$\chi^a(\vv{\kappa}_i \in {\kappa})$ to the form
$\chi^a(\vv{\kappa}_j) = \exp[2\pi i \ell_a (j - 1)/6] \chi^a(\vv{\kappa}_1)$, where $\ell_a$ is the angular momentum of spinor component $a$.
Using this fact, we can rewrite each component of $\psi_{{\kappa}, \ell_{{\kappa}}}$ (Eq.~\eqref{eq:three_patch_wavefunctions} as follows:

\begin{equation}
     \psi^a_{ \kappa,\ell_{ \kappa}}(\vv r) =   \varphi_{{\kappa},\ell_{\kappa} + \ell_a}(\vv r)\chi^a(\vv{\kappa}_1).
    \label{eq:basis-expanded-planewave}
\end{equation}
The spatial wavefunction is given by
\begin{equation}\label{eq:basis-real-space}
    \varphi_{{\kappa},  \ell_{\mathrm{tot}}}(\vv{r})\coloneqq\frac{1}{\sqrt{3}}\sum_{j=1,3,5} 
    e^{2\pi i \ell_{\mathrm{tot}} j/6}
    e^{i\vv \kappa_j\cdot \vv r},
\end{equation}
in which $\ell_{\mathrm{tot}}$ is the total angular momentum of the component. At $\kappa$ point, it can be written as $\ell_{\mathrm{tot}}=\ell_{\kappa}+\ell_a$. The functions $\varphi_{{\kappa}'}$ for ${\kappa}'$ can be obtained by complex conjugation.
Some of the key properties of these wavefunctions are reviewed in App.~\ref{app:real_space_properties}.

Let us now consider the RMG-inspired spinor, Eq.~\eqref{eq:three-path-spinor}, with $N_L=2$:
\begin{equation}
\label{eq:three-component-spinor}
    \chi^{(2)}(\vv\kappa_{n}) \propto (1, \eta e^{2\pi i(n-1)/6}).
\end{equation}
Since the components have different angular momenta $\ell_a$, their charge densities $\rho_{{\kappa}, \ell_\mathrm{tot}}(\vv{r}) = \sum_a |\psi_{{\kappa}, \ell_{{\kappa}}}^a (\vv{r})|^2$ appear offset from each other, resulting in the charge density pattern shown in 
the left column of Fig.~\ref{fig:3point-model-hartree}. We likewise show the charge densities in ${\kappa}'$ in the right column. 

The total charge densities are simply the sum of these charge densities. We note that for each Chern number, there are three possible charge densities corresponding to different $C_3$ angular momenta. In Sec.~\ref{sec:spotchecks}, we show that these schematic charge densities agree well with actual SCHF charge densities.

These spatial density patterns can be used to understand some aspects of the energetics of the three-patch states. In App.~\ref{app:three-patch-energy-competition}, we show that the three-patch charge densities can be used to understand 1) Hartree energy competition between $C=1$ and $C=0$ state, 2) Fock energy competition between $C=0$ and $C=1$ at small $\eta$, and 3) the stability of the $C=0$ state under honeycomb potential.

We caution the readers that even though the charge densities provide a simple graphical picture to understand some aspects of the energy competition, it also fails to predict many important features. In particular, it does not capture the strength of the Fock interaction at large $\eta$, thus failing to capture the stability of the AHC state.

\section{Application of the three-patch model to Rhombohedral Multilayer Graphene}
\label{sec:RMG}

We now introduce a streamlined microscopic model of rhombohedral multilayer graphene (RMG)~\cite{zhang_band_2010}.  We note that our model neglects detailed effects such as longer-range hoppings within the rhombohedral graphene, which was shown to not affect the phase diagram strongly~\cite{dong_anomalous_2023}, lattice relaxation,  and the layer-dependence of the Coulomb interactions. We also neglect the hBN moir\'e potential~\cite{jung_ab_2014,moon_electronic_2014,zhang_nearly_2019,park_topological_2023}, but allow for interaction generated spontaneous translation breaking. We show this simplified model hosts an AHC state within SCHF, and furthermore provides a concrete justification for the three-patch model and its approximations.

\subsection{Hamiltonian}

\begin{figure}
    \centering
    \includegraphics[width=\linewidth]{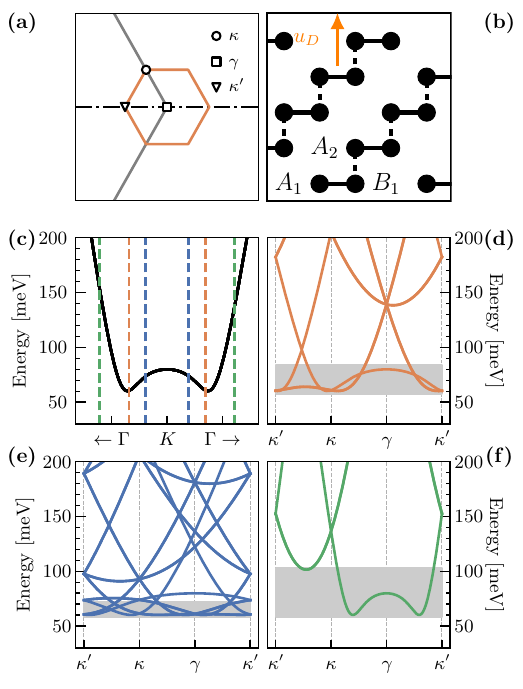}
\caption{(a) The choice of mBZ for folding. The ${\gamma}$ point is chosen to coincide with the $K$ point of RMG. (b) Schematic side view of the structure of RMG, showing the staircase-like structure. We simplify the RMG model such that only the hoppings $t_{0,1}$ along the staircase are taken into account. (c) The band structure of the simplified Hamiltonian of RMG at $N_L=5$ (Eq.~\eqref{eq:RMG_fullham}) along the $k_x$ axis at $u_D = \SI{40}{\milli\electronvolt}$.  Vertical lines correspond to the position of $\vv{\kappa}$ point after folding the band structure at densities $(29,9.2,2.9)\times 10^{11}\mathrm{cm}^{-2}$ from outside to inside. (d), (e), (f) The RMG band structure after folding according to (c). Electronic densities are $(9.2,2.9,29)\times 10^{11}\mathrm{cm}^{-2}$ respectively. Gray shading corresponds to Coulomb interaction strengths at the interparticle distance corresponding to these densities. Only the electronic density in (d) is suitable for the three-patch model. (e) is below the optimal range of density for the three-patch model due to the degeneracy at $\vv\gamma$. (f) is above the optimal range of density because the bandwidth is too large.}
    \label{fig:RMG_overview}
\end{figure}

Consider $N_L$ layers of graphene with rhombohedral stacking with creation operators $\hat{c}^\dagger_{\sigma,\ell}$ where $\sigma \in \st{A,B}$ labels sublattice and $\ell \in \st{1,\dots,N_L}$ labels layer. From the side (Fig.~\ref{fig:RMG_overview}(d)), RMG forms a staircase with different hopping strengths within and between layers --- akin to the SSH chain. The Hamiltonian is
\begin{equation}\label{eq:RMG_fullham}
\begin{aligned}
    \hat{h}_{\mathrm{RMG}} &= \sum_{\k} \hat{h}_0(\k) + \hat{h}_D(\k)\\
    \hat{h}_0(\k) &= -t_0 S(\vv k) \sum_{\ell=1}^{N_L} c^\dagger_{B,\ell}(\vv k)c_{A,\ell}(\vv{k})\\
    &\qquad + t_1 \sum_{\ell=2}^{N_L}c^\dagger_{B,\ell-1}(\vv{k})c_{A,\ell}(\vv{k}) +\mathrm{h.c.}\\
    \hat{h}_D(\k) &= -\sum_{\ell=1}^{N_L}u_D(\ell-1)\hat{n}_\ell(\vv k)
\end{aligned}
\end{equation}
where $t_0=\SI{3100}{meV}$ is the intralayer hopping, $t_1\approx \SI{380}{meV}$ is the interlayer hopping, $S(\vv k)=\sum_{n=0}^2 e^{i\vv k\cdot \vv \delta_n}$ in terms of graphene interatomic vectors $\vv \delta_n = C_3^{n-1} (0,a/\sqrt{3})$, $a=\SI{0.246}{nm}$ is the graphene lattice constant, and $\hat{n}_{\ell}(\vv k)=c^\dagger_{A,\ell}(\vv{k})c_{A,\ell}(\vv{k})+c^\dagger_{B,\ell}(\vv{k})c_{B,\ell}(\vv{k})$ is the electron density on layer $\ell$. Finally, $u_D>0$ is a displacement field that polarizes electrons in the conduction bands towards the top layer. We have neglected further-neighbor hoppings $t_{i\ge 2}$.  While they change the topology of the Fermi surface at small electron density from an annulus to three distinct pockets, a previous SCHF study has shown that the AHC state is unaffected by these further-neighbor hoppings~\cite{dong_anomalous_2023}. The model enjoys $C_3$ rotation symmetry, two-fold rotation $C_{2x}$, mirror $M_x$, time-reversal, and standard discrete translations.

We can define a natural length scale of this Hamiltonian by the condition that the first and second term of $\hat{h}_0(\vv{k})$ in Eq.~\eqref{eq:RMG_fullham} have equal magnitude. Approximating the intra-layer dispersion linearly as $t_0 S(K + \Delta \vv{k}) = \hbar v_F \Delta\vv{k} \cdot {\sigma}$ with $v_F = 1.004 \times 10^6 \si{\meter\per\second}$, we find the length scale to be $L_0 = h v_F / t_1 \approx 11 \si{\nano\meter}$. This natural length scale should be contrasted against the length scale of translation symmetry breaking, coming from either explicit moir\'{e} potential or spontaneous symmetry breaking. We expect qualitatively new physics to happen when these two length scales match. We note that this is precisely what happens for the magic angle of twisted bilayer graphene, whose moir\'{e} length scale is also around $10 \si{\nano\meter}$.

We consider an interacting Hamiltonian
\begin{equation}\label{eq:RMGintham}
    \hat{H}_{\mathrm{RMG}} = \hat{h}_{\mathrm{RMG}} + \frac{1}{2A} \sum_{\q} V_{\q} :\hat{\rho}_{\q} \hat{\rho}_{-\q}:, \ V_{\vv q} = \frac{2\pi \tanh |\vv q|d }{\epsilon_r \epsilon_0 |\vv q|}
\end{equation}
where $\hat{\rho}_{\q}$ is the charge density operator at wavevector $\q$, $A$ is the sample area, $V_{\vv q}$ corresponds to the double-gated Coulomb interactions with gate distance $d = \SI{25}{nm}$ and dielectric $\epsilon_r = 5$ (unless otherwise specified). Finally, normal ordering is with respect to the fermionic vacuum at charge neutrality for simplicity. We note this vacuum is strongly renormalized in realistic models in the limit $u_D \to 0$, necessitating ``Hartree-Fock subtraction" or other techniques to property renormalize the kinetic term in order to model the realistic system~\cite{kwan_moire_2023}. In the simplified model used here we take only the lowest conduction band to be dynamical.

To perform Hartree-Fock calculations, we focus on the low-energy part of the single-particle spectrum and only keep the states near the K point of graphene. We use graphene Bravais lattice vectors
\begin{equation}
    \vv{R}_1 = a(1, 0), \quad \vv{R}_2 = a(1/2, \sqrt{3}/2),
\end{equation}
with corresponding reciprocal lattice vectors
\begin{equation}
    \vv{G}_1 = \frac{2\pi}{a}(1, 1/\sqrt{3}), \quad \vv{G}_2 = \frac{2\pi}{a}(0, 2/\sqrt{3}),
\end{equation}
and $\vv{K} = 2\pi/a(2/3, 0)$. The vicinity of the K point is shown in Fig. \ref{fig:RMG_overview}(a).

To account for the possibility of translation symmetry breaking, we consider a ``ghost'' superlattice generated by the Bravais vectors
\begin{equation}
    \vv{r}_1 = L_s (1, 0), \quad \vv{r}_2 = L_s (1/2, \sqrt{3}/2),
\end{equation}
where $L_s$ is the superlattice scale defined below. The corresponding reciprocal lattice vectors are given by
\begin{equation}
    \label{eq:mini_reciprocal_lattice}
    \g_1 = \frac{2\pi}{L_s} (1,1/\sqrt{3}), \quad
    \g_2 = \frac{2\pi}{L_s} (0,2/\sqrt{3}).
\end{equation}
These generate the mini-Brillouin zone (mBZ) shown in Fig. \ref{fig:RMG_overview}(a). Here, we have chosen the mBZ to be centered at the graphene $K$ point and oriented the same way as the graphene Brillouin zone, with high-symmetry points ${\gamma}, {\kappa},$ and ${\kappa}'$.

The superlattice scale $L_s$ is fixed by the electron density $n$. As we are interested in the behavior of the system at the filling of one electron per superlattice unit cell i.e. $\nu=1$, we choose $L_s$ to enforce this filling. Since the area of the superlattice unit cell is $\sqrt{3}L_s^2/2$, we have
\begin{equation}
    L_s^2 = \frac{2}{\sqrt{3}n}.
\label{eq:density_versus_Ls}
\end{equation}

While we do not consider the effect of a moir\'{e} potential in this work, we provide a conversion between moir\'e parameters and superlattice parameters for ease of comparison to experiments with moir\'e patterns from proximate hBN layers. In the presence of moir\'{e} potential, the superlattice constant $L_s$ should be taken to be the same as moir\'{e} lattice constant. In the case of the moir\'{e} pattern generated by hBN/graphene lattice mismatch, the lattice constant goes as

\begin{equation}
    L_s^{\text{moir\'e}} = a \frac{1 + \epsilon}{\sqrt{\epsilon^2 + 2(1+\epsilon)(1-\cos\theta)}},
\end{equation}
where $\epsilon\approx 0.018$ is the lattice mismatch, and $\theta$ is the relative rotation angle between hBN and graphene. The experimental value of $\theta=0.77^\circ$~\cite{lu_fractional_2024} corresponds to $L_s \approx 11.5 $nm, putting it close to the natural length scale $L_0$. Plots of this relation are shown in Appendix~\ref{app:gate_dependence}.

With the above choice of geometry, the single particle states in our mean-field calculations take the form
\begin{equation}
    \psi_{\vv{k}}(\vv{r}) = \sum_{m, n} c_{\vv{k}}^{m,n}\phi_{\vv{k} + m \vv{g}_1 + n \vv{g}_2}(\vv{r}),
\end{equation}
where $\phi_{\vv{k}}$ are eigenstates of the single particle Hamiltonian, and $c_{\vv{k}}^{m,n}$ are complex coefficients.

\begin{figure*}
    \centering
    \includegraphics[width = .99\textwidth]{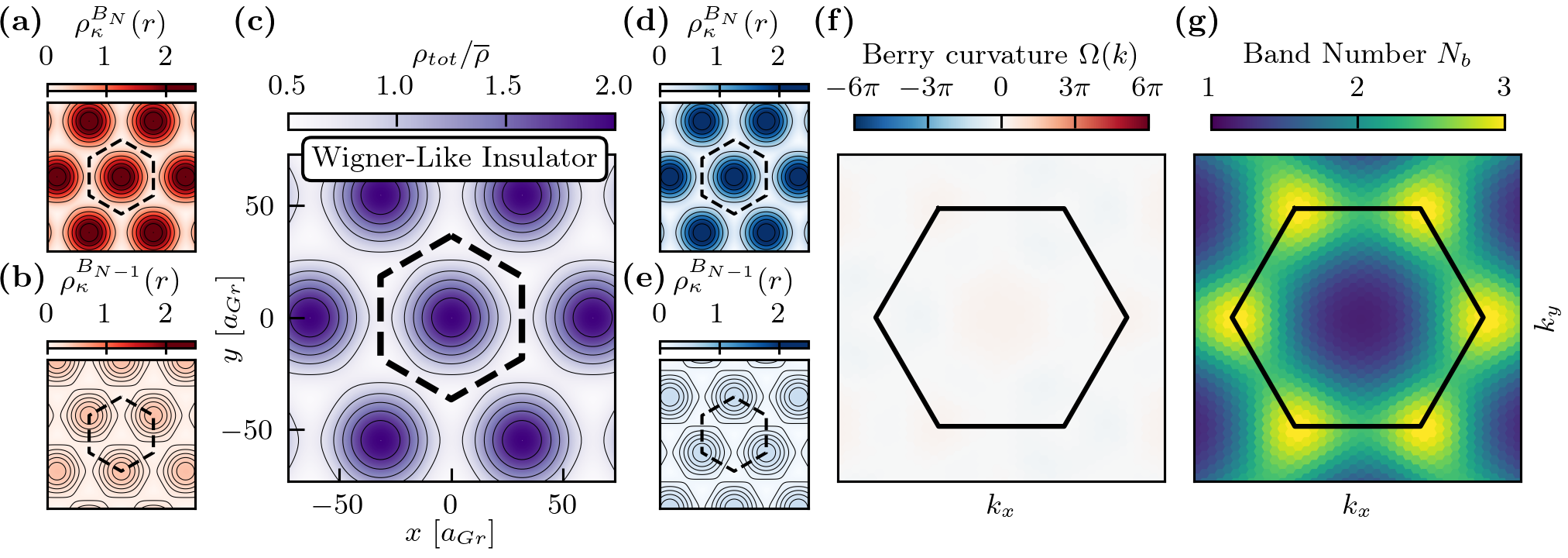}
    \caption{Properties of a $C=0$ state (Wigner-like insulator) found in SCHF at $N_L=5$ with parameters $(t_0, t_1) = (3100, 380)\si{meV}$, $u_D = 15\si{meV}$, $n = 4.8\times 10^{11}\si{cm^{-2}}$, analogous to that in Fig.~\ref{fig:C1_spotcheck}. We use $24\times 24$ $k$-points and project onto the lowest $7$ bands.
    (a-e) Charge densities of the $C=0$ state. The left panels (a,b) correspond to orbitals $B_{N_L}$ and $B_{N_L-1}$ of the wavefunction at ${\kappa}$, while the right panels (d,e) correspond to charge densities at ${\kappa}'$ of the same orbitals. The central panel (c) is the total charge density, including all the momenta points.  (f) The Berry curvature, normalized such that the average of $\Omega(\vv{k})$ being $2\pi$ corresponds to $C=1$ over the BZ.(g) The band number $N_{\textrm{b}}(\vv{k})$ in the mBZ. We can clearly see the existence of three patches. 
    }
    \label{fig:C0_spotcheck}
\end{figure*}

\subsection{Spinor structure of RMG}

We now investigate the spinor structure of RMG. As has been shown in Sec.~\ref{sec:three-patch-model}, a fundamental input to the three-patch model are the spinors at the corners of the mBZ, i.e. the $\vv\kappa_i$ points.

We assume the time-reversal symmetry has been broken by spontaneous valley polarization, and focus on near the graphene $K$ point. We can simplify Eq.~\eqref{eq:RMG_fullham} by approximating $-t_0S(\vv K + \vv k) \approx v_F \bar{k}$, where $v_F \approx 10^{6} \si{m/s}$ is the graphene Fermi velocity, and $\bar{k} = k_x-ik_y$. When $v_F |\vv k|<t_1$, the spectrum corresponds to the topological phase of the SSH chain~\cite{xiao_density_2011,su_solitons_1979}. It hosts two localized boundary modes, polarized on A and B sublattices. The $A$ mode has (unnormalized) wavefunction
\begin{equation}\label{eq:spinor}
    \tilde{\chi}_B(\vv K + \vv k) = (1, 0, -\eta(\vv k), 0, \eta(\vv k)^2 \dots) \text{ where } \eta(\vv k) = \frac{v_F |\vv k|}{t_1}
\end{equation}
in the basis $B_{N_L},A_{N_L},B_{N_L-1},A_{N_L-1},\dots$. Its normalized counterpart is $\chi_B =\tilde{\chi}_B/|\tilde{\chi}_B|$. The $A$ boundary mode is $\chi_A = \sigma_x P \chi_B^*$ where $\sigma_x$ flips sublattice, and $P$ reverses the order of the layers.

If $u_D, v_F|\vv k| < t_1$, then the layer edge modes give the that give the highest energy valence band and lowest energy conduction band of RMG. We can thus project the single-particle Hamiltonian $H(\vv k)$ into a two-dimensional Hilbert space spanned by $\chi_A(\vv{k})$ and $\chi_B(\vv{k})$:
\begin{equation}\label{eq:effham}
    H^{\textrm{eff}}(\vv K + \vv k) = \begin{pmatrix}
        -u_D \delta(\vv k) & t_1 \eta(\vv k)^{N_L}\\
        t_1 \eta(\vv k)^{*N_L} & u_D (-(N_L-1)+\delta(\vv k))
    \end{pmatrix}
\end{equation}
where $\delta(\vv k)=\frac{1}{u_D}\braket{\chi_A(\vv{k})| H_D | \chi_A(\vv{k})}\approx|\eta(\vv k)|^2/(1- |\eta(\vv k)|^2)$. At small enough $\vv{k}$, the diagonal part dominates, and we can approximate the energy eigenstate by $\tilde{\chi}_B$.

Dropping the $A$ sublattices, and performing a layer-dependent gauge transformation, we find that the spinor at the $\vv{k}$ are given by
\begin{equation}\label{eq:k_spinors}
    \chi_B(\vv K + \vv k) \propto (1,\eta(\vv{k}),\eta(\vv{k})^2,\dots).
\end{equation}
Thus, we recover the spinor structure used in Sec. \ref{subsec:three-patch-RMG}, allowing us to compare the phenomenology of the section to that of the simplified RMG Hamiltonian.

We now consider the mBZ at density $n$ to determine the value of $\eta$. The mBZ corners, $\kappa_j$, measured from the graphene K point, are given by

\begin{equation}
    \vv\kappa_j(n) = \frac{2\pi}{L_s(n)}(C_6)^{j-1}(2/3, 0),
\end{equation}
where $C_6$ is the six fold counterclockwise rotation, and $L_s$ is determined by $n$ via Eq.~\eqref{eq:density_versus_Ls}. Correspondingly, the spinors are given by

\begin{equation}\label{eq:kappaspinors}
    \chi_B(\vv K + \vv\kappa_{n+1}) \propto (1,\eta(\vv{\kappa}_i(n)) e^{2\pi in/6},\eta(\vv{\kappa}_i(n))^2 e^{4\pi in/6},\dots)
\end{equation}
where
\begin{equation}
    \eta(\vv{\kappa}_i(n))\coloneqq \frac{v_F|\vv\kappa_i(n)|}{t_1} =\frac{L_0|\vv\kappa_i(n)|}{2\pi} = \frac{L_0}{2\pi^2} \left(\frac{n}{6\sqrt{3}} \right)^{\frac{1}{2}}.
\label{eq:eta_from_density}
\end{equation}
The critical $\eta\sim0.6$ in the three-patch model, where the $C=0$ state transitions to a $C=1$ state, corresponds to an electronic density of $7.9\times 10^{11}\,\si{cm^{-2}}$, with a corresponding superlattice length scale of approximately $\SI{12}{nm}$, slightly larger than the length scale $2\pi v_F / t_1 = \SI{11}{nm}$. In terms of the angle between hBN and graphene, this corresponds to roughly $0.6^\circ$, close to the experimental value $0.77^\circ$ \cite{lu_fractional_2024}.

\subsection{Density range of validity}
    
\begin{figure*}
    \centering
    \includegraphics[width = .99\textwidth]{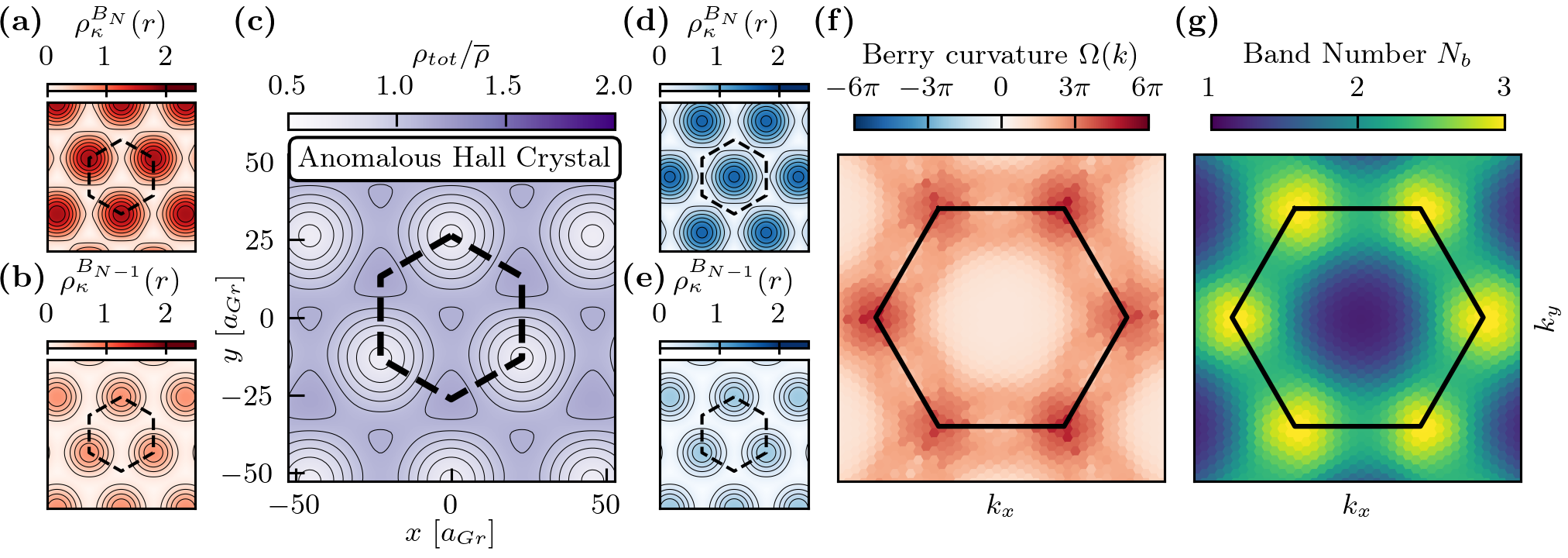}
    \caption{Properties of a $C=1$ state (anomalous Hall crystal) found in SCHF at parameters $N_L=5$, $(t_0, t_1) = (3100, 380)\si{meV}$, $u_D = 40\si{meV}$, $n = 9.2\times 10^{11}\si{cm^{-2}}$. Panels are arranged similarly to Fig.~\ref{fig:C0_spotcheck}.}
    \label{fig:C1_spotcheck}
\end{figure*}

The band structure of RMG under displacement field, Fig.~\ref{fig:RMG_overview}(c), has a relatively flat band bottom, suggesting it could host three-patch states. We now carefully demarcate the density regime where expect the three-patch phenomenology.

We then fold the band structure onto the mini-Brillouin zone given according to the reciprocal lattice vectors in Eq.~\eqref{eq:mini_reciprocal_lattice}. Note that since there is no superlattice-scale potential, there is no eigenvalue repulsion.

There is a range of densities in which the minimal model is applicable.
The dashed lines in Fig.~\ref{fig:RMG_overview}(c) show the location mBZ boundaries at $k-K = 
\pm 2\pi/L_s(2/3, 0)$ corresponding to three different densities. Correspondingly, the xaxis labels ($\kappa'$, $\kappa$, and $\gamma$) on Fig.~\ref{fig:RMG_overview}(d-f) refer to different points for different panels.
Fig.~\ref{fig:RMG_overview}(d)
shows the band structure at optimal density (orange line), whose lowest band is relatively flat compared to the interaction scale (gray shading), promoting symmetry breaking.
Moreover, the large gap to the second band exceeds the interaction scale significantly at $\gamma$, disfavoring mixing. Thus, the ground state is likely to be a weakly coupled symmetry-broken state.

Outside this range of densities, the three-patch assumptions can break down. If the density is too low [Fig. \ref{fig:RMG_overview}(e)], then the other bands will mix strongly at $\gamma$, violating the assumed form of the wavefunction in Eq.~\eqref{eq:three_patch_wavefunctions}. This will potentially alter the angular momentum $\ell_\gamma$. Alternatively, too large a density [Fig. \ref{fig:RMG_overview}(f)] makes the band too dispersive at the mBZ boundaries, potentially favoring a Fermi liquid over a crystal.
Henceforth we focus on the ``Goldilocks zone" of densities corresponding to folding the bands near the edge of the flat band bottom.


\subsection{Three-patch states as SCHF ground states}
\label{sec:spotchecks}

In Sec.~\ref{sec:three-patch-model}, we showed that there is a close energetic competition between $C=0$ and $C=1$ states in the three-patch model using RMG spinors. We now show that the microscopic model Eq.~\eqref{eq:RMGintham} not only realizes this energetic competition, but recapitulates the same real space charge density patterns.

We briefly comment on our SCHF numerics. We explicitly allow translation symmetry breaking by using the folded band structure in the mBZ defined in Eq.~(\ref{eq:mini_reciprocal_lattice}). We assume valley and spin polarization, resulting from the flavor Stoner ferromagnetism due to the large density of states~\cite{zhou_half-_2021}. We choose the superlattice scale $L_s$ such that the filling is one electron per superlattice unit cell at each density considered. We project to the lowest 7 conduction bands for simplicity,  although we note that it may become uncontrolled due to the reduced Hilbert space. See Ref.~\cite{dong_anomalous_2023} for further numerical details.

At low densities, corresponding to low $\eta$, the ground state is a topologically trivial insulator. Its real-space charge density is shown in Fig.~\ref{fig:C0_spotcheck}(c), which has an overall triangular lattice of a Wigner-like insulator --- just as in Fig.~\ref{fig:3point-model-hartree}. This charge density is essentially determined by the sum of charge densities from ${\kappa}$ and ${\kappa}'$.
In the dominant $B_{N_L}$ spinor components, shown in (a) and (d), the charge is centered at the same Wyckoff positions, with the subdominant $B_{N_L-1}$ components ``filling in the gaps" to produce the same structure shown in Fig.~\ref{fig:3point-model-hartree_app}. The Berry curvature, shown in Fig.~\ref{fig:C0_spotcheck}(f) is small everywhere, giving $C=0$. 

At higher densities, i.e. higher $\eta$, there is a transition to a $C=1$ state. Fig.~\ref{fig:C1_spotcheck}(c) shows its charge density, which is fairly uniform with peaks on a honeycomb lattice. Now the dominant $B_{N_L}$ components are centered on \textit{different} Wyckoff positions at $\kappa$ versus $\kappa'$, but the subdominant components are on top of each other. Referring to Fig.~\ref{fig:3point-model-hartree}, we can see this is precisely the same structure as the $C=1$ state there, whose dominant and subdominant components are represented by large and small circles, respectively. Here the Berry curvature, Fig.~\ref{fig:C1_spotcheck}(f), is strongly peaked at $\kappa$ and $\kappa'$, with a slight $C_2$ breaking. 

The success of the three-patch model, which only contains three single-particle wavefunctions, to reproduce the SCHF charge density is consistent with three-patch assumptions. We now explicitly verify that the ground states are weak-coupling states that mix only a few low-lying bands in the folded band structure: this can be revealed by studying the number of single-particle bands of the folded band structure mixed into the occupied band of the SCHF ground state. To quantify this, we note that SCHF eigenstates have the form
\begin{equation}
    \psi_{\vv k}(\vv r) = \sum_{\vv g} \chi_{\vv k}(\vv g) e^{i (\vv k+\vv g)\cdot\vv r},
\end{equation}
where $\vv g$ runs over superlattice reciprocal vectors, i.e. different (unhybridized) mini-bands. We define a mini-band distribution function
\begin{equation}
    p_{\vv k}(\vv g) \coloneqq |\chi_{\vv k}(\vv g)|^2.
\end{equation}
When the eigenstate is an equal superposition of $N$ plane waves (i.e. $N$ mini-bands after folding), the von Neumann entropy of the band distribution function $S_{\mathrm{v.N.}}(\vv k) = -\sum_{\vv g}p_{\vv k}(\vv g) \log p_{\vv k}(\vv g)$ equals $\log N$. Therefore, we define the band number
\begin{equation}
    N_b(\vv k) \coloneqq e^{S_{\mathrm{v.N.}}(\vv k)}=\exp\left(-\sum_{\vv g}p_{\vv k}(\vv g) \log p_{\vv k}(\vv g)\right)
\end{equation}
 to quantify the number of bands that mix at each $k$-point.

The band number $N_b(\vv k)$ is shown in Fig.~\ref{fig:C0_spotcheck},\ref{fig:C1_spotcheck} (g). It is approximately $1$ near ${\gamma}$, but is close to 3 in at ${\kappa}, {\kappa}'$. This confirms that only minibands close to the band bottom appear in the SCHF wavefunction, explicitly confirming the ``weak coupling" assumption.
In fact, we can see the appearance of large patch-like features near $\gamma$, ${\kappa}$ and $ {\kappa}'$. The assumptions of the three-patch model are therefore borne out in microscopic SCHF, so we should expect the phenomenology and predictions of the three-patch model to hold.

\begin{figure}
    \centering
    \includegraphics{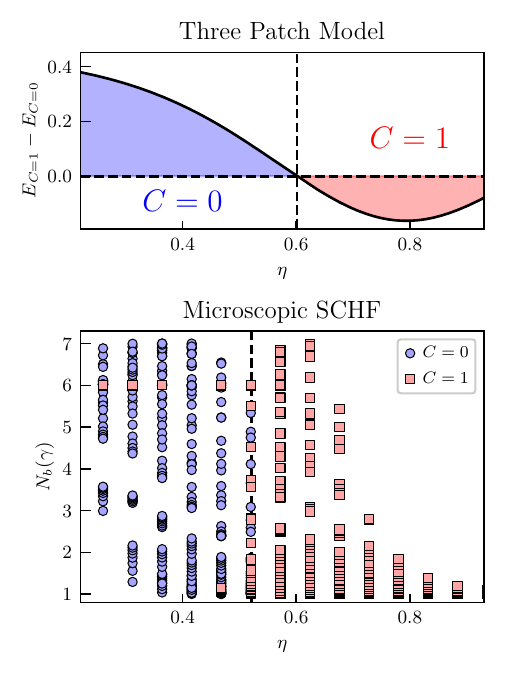}
    \caption{
    (a) Energy differences between $C=1$ and $C=0$ ground states in the three-patch model. We take spinors of the form Eq. \eqref{eq:three-path-spinor} with $N_L =5$, finding a transition at $\eta \approx 0.6$. 
    (b) Dimensionally-reduced phase diagram of insulating states in Eq.~\eqref{eq:RMGintham}, simplified microscopic RMG. Here $\eta$ is proportional to the density, while $N_b(\gamma)$ quantifies how much remote bands are populated in the $\gamma$ point wavefunction of the SCHF ground state (details in text). The line at $\eta=0.52$ approximately separates the $C=0$ and $C=1$ phases, in good agreement with the phenomenological three-patch model. 
    }
    \label{fig:schematic_phase_diagram}
\end{figure}

\subsection{Microscopic verification of the three-patch phase diagram}

The previous sections demonstrated translation breaking, weak band mixing and clear patches in SCHF ground states of RMG, putting the model into the regime of the three-patch model. We therefore expect the predictions and phenomenology of the three-patch model to appear in the microscopic model. We now show a striking manifestation of this: the competition between $C=0$ and $C=1$ is set by a single parameter $\eta$.

Fig.~\ref{fig:schematic_phase_diagram}(a) shows the energy difference between $C=0$ state and the $C=1$ state in the three patch model with Coulomb interaction using spinors $\chi_B$ from Eq.~\eqref{eq:spinor} parametrized by $\eta$ with $N_L=5$. There is a phase transition around $\eta=0.6$. We recall that in RMG, the value of $\eta$ is given by 

\begin{equation}
    \eta(n) = \frac{L_0}{2\pi^2} \left(\frac{n}{6\sqrt{3}} \right)^{\frac{1}{2}}
\end{equation}
where $L_0 \approx 11 nm$ is the natural length scale of RMG.
We note that effect of the ignored zero mode $\chi_A$ is negligible at large $u_D$, where its weight is suppressed. However, at small $u_D$, its weight can become as large as $O(10) \%$. This gives rise to a large ``gray area'' where the three-patch model is inconclusive about the ground state (App.~\ref{app:realistic_spinors}), and effects neglected in the three-patch model can determine the ground state. For simplicity, we apply the result of the three-patch model with $\chi_B$ in our analysis below.

If $\eta$ is the primary driver of the phase transition between $C=0$ and $C=1$ --- as predicted by the three-patch model --- there should be a clean phase boundary in terms of $\eta$ alone. To verify this, we first compute the microscopic ground state over a high dimensional parameter space, then dimensionally reduce the results in terms of just two variables. Explicitly, we select $\eta$ and $N_\textrm{b}(\gamma)$, which quantifies the number of remote bands that mix at the $\gamma$ point, and must be close to $1$ for the three patch model to be applicable. Concretely, we take $N_L=5$, and vary $u_D\in[5,70]\,\si{meV}$,   $\epsilon_r\in[3,10]$, and $n\in[1.4,17]\times 10^{11}\si{cm}^{-2}$. We identify metallic data points either by a large occupation difference in the mBZ or by a small indirect gap less than $\SI{2}{meV}$. We discard such metallic data points as they are outside the scope of the three-patch model.

The dimensionally-reduced SCHF phase diagram is shown in Fig.~\ref{fig:schematic_phase_diagram}(b). We see that all of the points above $\eta=0.52$ (dashed line) belong to the $C=1$ phase, while most of the points below $\eta=0.52$ are in the $C=0$ phase. We note that the $C=0$ and $C=1$ phase boundary in SCHF at $\eta = 0.52$ goes far beyond the weak coupling $N_b(\gamma) = 1$ point, potentially hinting that the three-patch model works beyond the weak-coupling assumption. We may thus conclude that $\eta$ indeed controls the topological phase transition, and that the three-patch model accurately predicts the critical value.

\begin{figure}
    \centering
    \includegraphics{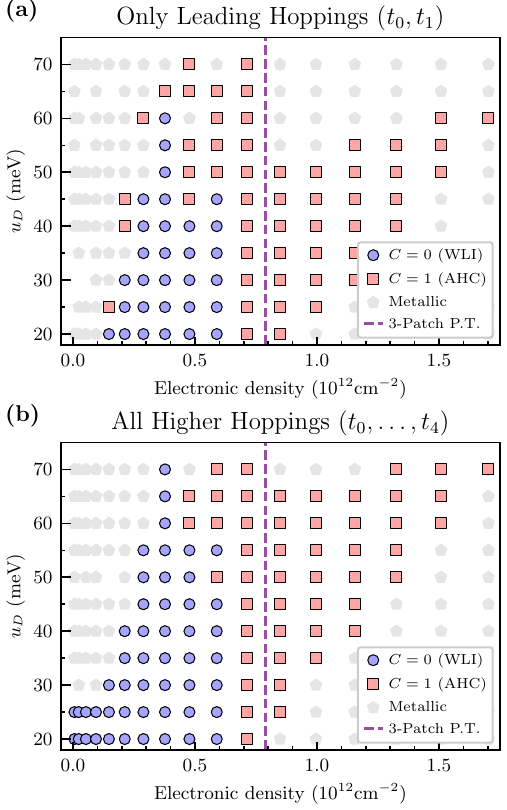}
    \caption{Phase diagram of RMG without a moir\'e potential in self-consistent Hartree-Fock. Red square indicate anomalous Hall crystals, blue circles indicate Wigner-crystal-like insulating states. Gray pentagons are states that are likely metallic (diagnosed by $n(\vv k)$ varying by $>1\%$). The dashed purple line shows the phase transition predicted by the three-patch model --- close to the true mean field transition. (a) Phase diagram in simplified model Eq.~\eqref{eq:effham}. (b) Phase diagram including higher hoppings $(t_2,t_3,t_4) = (\SI{-21}{meV}, \SI{290}{meV}, \SI{141}{meV})$ in the expanded model from Ref.~\cite{dong_anomalous_2023}. We note that the addition of higher hoppings does not significantly alter the phase diagram, suggesting these phase are stable to a variety of perturbations.
    Parameters: $24\times 24$ unit cells, $\epsilon_r = 5$, $d_{\mathrm{gate}} = \SI{10}{nm}$.
    }
    \label{fig:density_u_D_phase_diagram}
\end{figure}

It is instructive to fix a realistic value of $\epsilon_r$ to see a density v. $u_D$ phase diagram(Fig.~\ref{fig:density_u_D_phase_diagram}(a)). We see that the AHC state is favored in the ``Goldilocks'' region at intermediate density and a narrow range of displacement field $u_D$. The qualitative features of this phase digram remains the same even when we include further-neighbor hoppings(Fig.~\ref{fig:density_u_D_phase_diagram}(b)). As predicted by the three-patch model, the AHC phase gives way to the WC phase at low density.
In fact, the critical value can be shifted to lower density/$\eta$ by changing the gate distance of the interaction, exactly as predicted by the three-patch model (App.~\ref{app:gate_dependence}). We will comment further on the striking success of the three-patch model in the discussion.

\subsection{Three-patch states from analytical mean-field equation}\label{sec:gap_equation}

In this brief section, we use a mean field gap equation to analyze the interaction-driven gap. Surprisingly, we find gap opening induces nearly constant wavefunctions in the $\kappa, \kappa'$ patches.

Let us focus on a small momentum patch of radius $S$ at $\kappa$. At each point in the patch, there are three low-lying states in the Hilbert space of the folded Brillouin zone corresponding to plane waves $\ket{\phi_{\vv{\kappa}_1 + \vv{k}}}$, $\ket{\phi_{\vv{\kappa}_3 + \vv{k}}}$, and $\ket{\phi_{\vv{\kappa}_5 + \vv{k}}}$, defined in Eq.~\eqref{eq:singleparticlestates}. In this three-dimensional Hilbert space, the linearized dispersion of each plane-wave state (before folding) is given by

\begin{equation}\label{eq:1pham_gap}
    H_0(\vv{k}) =
    \begin{pmatrix}
        v_F^{\textrm{eff}} \vv{k} \cdot \hat{\vv{e}}_1 & 0 & 0 \\
        0 & v_F^{\textrm{eff}} \vv{k} \cdot \hat{\vv{e}}_3 & 0 \\
        0 & 0 & v_F^{\textrm{eff}} \vv{k} \cdot \hat{\vv{e}}_5
    \end{pmatrix},
\end{equation}
where $\hat{\vv{e}}_i$ is the unit vector parallel to $\vv{\kappa}_i$, and $v_F^{\textrm{eff}}$ is the effective Fermi velocity at the $\kappa$ patch. We assume the dispersion $v_F^{\textrm{eff}}$ is small, since the $\kappa$ patch is near a local minimum of dispersion in experimental relevant regimes [see Fig. \ref{fig:RMG_overview}(c)].

We consider a mean-field Hamiltonian $H_{\textrm{mf}}(\vv{k})$ by projecting the Hatree and Fock terms into this subspace. By $C_3$ symmetry, it takes the form
\begin{equation}\label{eq:mfham_gap}
    H_{\textrm{mf}}(\vv{k})=
    \begin{pmatrix}
        0 & \Delta(\vv k) & \Delta^*(\vv k) \\
        \Delta^*(\vv k) & 0 & \Delta(\vv k) \\
        \Delta(\vv k) & \Delta^*(\vv k) & 0
    \end{pmatrix}.
\end{equation}
where the mean field $\Delta(\vv{\vv k})$ can be written as
\begin{align} 
\label{eq:meanfield}
    \nonumber
    &\Delta(\vv{\vv k}) = V_{\vv g_1}\Lambda_{\vv \kappa_1+\vv k,\vv \kappa_3+\vv k}\int d^2\vv k'\Lambda_{\vv \kappa_3+\vv k',\vv \kappa_1+\vv k'}\braket{\hat{c}^\dagger_{\vv\kappa_3+\vv k'}\hat{c}_{\vv\kappa_1+\vv k'}}\\
    &\quad -\int d^2\vv k' V_{\vv k-\vv k'}\Lambda_{\vv\kappa_3+\vv k',\vv\kappa_3+\vv k}\Lambda_{\vv\kappa_1+\vv k,\vv\kappa_1+\vv k'}\braket{\hat{c}^\dagger_{\vv\kappa_3+\vv k'}\hat{c}_{\vv\kappa_1+\vv k'}},
\end{align}
where $\Lambda_{\vv k,\vv k'}=\chi^\dagger(\vv k)\chi(\vv k')$ is the form factor of plane wave states. The first term is the Hartree contribution, whereas the second term is the Fock contribution.

Since we are only interested in the vicinity of ${\kappa}$, we  assume that all the spinors around the $\vv \kappa_i$ points are roughly unchanging:
\begin{equation}
    \chi(\vv \kappa_i+\vv k)\approx \chi_{i}.
\end{equation}
We also approximate $V_{\vv k-\vv k'}=V_{\vv 0}$. With these approximations, Eq.~\eqref{eq:meanfield} can be drastically simplified:
\begin{equation}
    \label{eq:delta_self_consistent}
    \Delta(\vv{\vv k}) = (V_{\vv g_1}|\Lambda_{1,3}|^2-V_{\vv 0})\int_{|\vv k'|<S}\braket{\hat{c}^\dagger_{\vv\kappa_3+\vv k'}\hat{c}_{\vv\kappa_1+\vv k'}}\equiv\Delta.
\end{equation}
Note that now the mean-field potential $\Delta$ is independent of $\vv k$, and thus so is the mean-field Hamiltonian Eq.~\eqref{eq:meanfield}.

To solve the self-consistent equation Eq.~\eqref{eq:delta_self_consistent}, we consider the single-particle Hamiltonian $H(\vv k)=H_0(\vv k)+H_{\textrm{mf}}(\vv k)$, and solve for the lowest energy single-particle state of $H(\vv k)$. In the absence of the kinetic part $H_0(\vv k)$, the lowest energy eigenstate takes the form
\begin{equation}
\label{eq:gap_equation_k_zero}
    \ket{\vv k^{(0)}} = \frac{1}{\sqrt{3}}(\ket{\phi_{\vv{\kappa}_1 + \vv k}} +\ket{\phi_{\vv{\kappa}_3 + \vv k}} +\ket{\phi_{\vv{\kappa}_5  + \vv k}}),
\end{equation}
if we take $\Delta$ to be real and negative. We note that the eigenstate at $\vv{k}=0$ has angular momentum $\ell_\kappa = 0$.
Now we treat $H_0(\vv k)$ perturbatively, assuming $\varepsilon=|v_F^{\textrm{eff}}\vv{k}/\Delta|$ to be small. The lowest energy eigenstate of the full Hamiltonian $H(\vv k)$ is given by
\begin{equation}
    \label{eq:gap_equation_components}
    \ket{\vv k} = \frac{1}{\sqrt{3}}\sum_{i=1,3,5}\left(1- \frac{(\hat{\vv k}\cdot \hat{\vv e}_i)}{3|\Delta|}\varepsilon\right)\ket{\phi_{\vv\kappa_i+\vv k}}+O\left(\varepsilon^2\right).
\end{equation}
Evaluating $\langle\hat{c}_{\boldsymbol{\kappa}_3+\boldsymbol{k}^{\prime}}^{\dagger} \hat{c}_{\boldsymbol{\kappa}_1+\boldsymbol{k}^{\prime}}\rangle$ with $\ket{\vv k}$, we find
\begin{equation}
\label{eq:gap_equation}
    \Delta= \left(V_{\vv 0}-V_{\vv g_1}|\Lambda_{1,3}|^2\right)\int d^2\vv k'\frac{\sqrt{2}\Delta}{\sqrt{|v_F^{\textrm{eff}}\vv k'|^2+18|\Delta|^2}}.
\end{equation}
Cancelling $\Delta$ from each side, it is easy to see the equation admits a nonzero solution only if the prefactor is positive.

The positivity condition on the prefactor has a quantum geometric interpretation. It can be written as
\begin{equation}
    V_{\vv 0}-V_{\vv g_1}|\Lambda_{1,3}|^2 = V_{\vv 0}-V_{\vv g_1} + V_{\vv g_1}|d_{1,3}|^2 > 0,
\end{equation}
where $|d_{1,3}|^2 = 1 - |\Lambda_{1,3}|^2$ is the quantum distance~\cite{bohm2003geometric}. This inequality is satisfied for physical interactions $V_{\vv g_1}<V_{\vv 0}$, revealing a mean field tendency to open a gap.
In fact, in the limit of the contact interaction $V_{\vv{g}_1} = V_{\vv 0}$, the left-hand side of the inequality reduces to $V_{{\vv g}_1}|d_{1,3}|^2$, revealing that the quantum distance between spinors is a major driving force behind the symmetry breaking tendency. We also note that the positivity of the prefactor is a necessary, but not sufficient, condition for opening a finite gap. Unlike in BCS theory, the $k$-integral does not diverge at $\Delta=0$, and thus the equation does not hold for an arbitrarily small prefactor. It is amusing to note that the magnitude of the form factors controls the gap, while their phase controls the topology of a resulting insulator.

From Eq.~\eqref{eq:gap_equation_components}, we see that the wavefunctions $\ket{\vv{k}}$ are all perturbatively close to $\ket{\vv{k}^{(0)}}$, controlled by the small ratio $|v_F^\mathrm{eff}\vv{k}/\Delta|$, giving rise to a patch-like behavior. 
The physical interpretation of this is straightforward: the Fock energy favors a large overlap between wavefunctions, similar to how it favors ferromagnetic configuration between spins with large, negative Fock energy. The states around $\kappa$ are therefore perturbatively close to each other, controlled by the dispersion $v_F^\mathrm{eff}$ that is small around the flat band bottom.

\section{Conclusion and outlook}
\label{sec:discussion}

In this work, we constructed a physically-motivated minimal ``three-patch model" for the AHC state composed of seven spinor-valued plane waves at the high symmetry points $\kappa$, $\kappa'$, and $\gamma$. In the three-patch model, these high symmetry point wavefunctions completely determine both the energetics and the topology of the states, through which we predict the phase diagram of the RMG.
The success of the phenomenological three-patch model is striking. Despite its extreme simplicity --- with analytic solvability, no kinetic energy, and a vastly reduced Hilbert space --- it qualitatively and quantitatively captures the results of microscopic SCHF calculations. This has a fascinating implication: the competition between $C=0$ and $C=1$ is mostly or fully determined by interactions.

Given the example of Wigner crystals, one might expect the topologically trivial electronic crystal to win the energetic competition. But here a crucial extra ingredient is present: the spinor structure. The spinors enter via two geometrical phases, $\theta_H$ and $\theta_F$, that determine the ground state favored by the Hartree and Fock terms respectively. These can compete, but there is a broad ``frustration free" regime where Hartree and Fock terms collaborate to favor $C=1$. In fact, the spinors of RMG fall in this ``frustration free'' regime in the broad range of the phase diagram.

When comparing careful self-consistent Hartree-Fock results on the simplified RMG Hamiltonian and Coulomb interaction with $N_L = 5$, we found that the three-patch model successfully described many features of RMG, at least in the weak-coupling regime. In particular, we found that the three-patch model qualitatively reproduced the phase diagram as shown in Fig.~\ref{fig:schematic_phase_diagram}. The phase boundary between the topological and trivial state was mostly determined by a single parameter $\eta$ of the spinor. Rather surprisingly, the three-patch model also captured the precise charge density pattern obtained from numerics, which in part justifies the effectiveness of our three-patch model.

It is fruitful to consider the unusual mechanism that generated the Berry curvature in the AHC state. Explicitly, the Berry curvature comes from phase-ful superpositions of plane waves produced from gap opening --- \textit{not} just the intrinsic Berry curvature present in the single particle bands before folding. Even if there is only an infinitesimal amount of intrinsic Berry curvature, such as for the $N_L=3$ spinor at $\eta \to 0$ with contact interactions (Fig.~\ref{fig:three-patch_hartree-fock_comparison}), the AHC state can still be favored. The AHC thus represents a novel scenario where interaction-driven spontaneous translation breaking is the \textit{source} of topology.
The high density of states regimes without much Berry curvature, which are often realized near the bottom of a band, may thus permit AHC states.
We suggest this mechanism opens an unexplored regime where quantized anomalous Hall effects might occur.

In fact, the three-patch model provides a simple algorithm to look for AHC candidate materials: 1) Perform \textit{ab initio} calculations to look for compound with the band structure with a flat band bottom, 2) Compute the spinor structure near the band bottom, and 3) Compute the quantum geometric phases $\theta_H$ and $\theta_F$. If these phases lie in the ``frustration free'' region, the compound is a potential AHC candidate.

While we formulated the three-patch model for continous symmetry breaking, a similar treatment based on symmetry indicators and the patch model might be applicable in more generality. For example, the appearance of QAH-crystal states in moir\'{e} systems~\cite{sheng2024quantumanomaloushallcrystal} can be subject to three-patch style analysis. It is also interesting to note that the combination of a weak translation-breaking potential and commensurate filling might stabilize AHC with other Bravais lattice geometries.

We note a simple experimental prediction of the three-patch model. Since the critical value of $\eta$ is proportional to $\sqrt{n}/t_1$, increasing the interlayer hopping $t_1$ by applying pressure would increase the density necessary to stabilize the AHC state. Pressure therefore might drive a quantum phase transition between the AHC state and the Wigner-like insulator state.

An essential future direction is to go beyond mean-field theory. Although the three-patch model here has given a simple physical picture of the AHC phase, we expect that --- just as for the Wigner crystal~\cite{goerbig2004competition, knoester2016electron, dora2023competition} --- its energetic competition should change significantly in beyond mean-field treatments. 
This would shed light on the energy competition with metallic states, as well as correlated insulators e.g. an analogue of composite fermion crystals in the lowest Landau level \cite{yi1998laughlinjastrow, chang2005microscopic, archer2013competing,zuo2020interplay}.
Moreover, there is a fascinating possibility that the AHC state can compete with a \textit{fractional anomalous Hall crystal}, where a larger unit cell is fractionally occupied, resulting in a translation broken topologically ordered state.
Understanding these questions requires analytical or numerical techniques that can treat many strongly-correlated bands, and will be a topic for future work. We hope that the three-patch model will provide valuable insight and phenomenological guidance for such future studies.

\noindent \textit{Note added}---After the completion of this work, Refs.~\cite{zeng_sublattice_2024,tan_parent_2024,dong_stability_2024} appeared. Refs.~\cite{zeng_sublattice_2024,tan_parent_2024} proposed different systems to realize the AHC. Ref.~\cite{dong_stability_2024} discussed a model similar to ours and the results agree in overlapping areas.

\begin{acknowledgements}
We acknowledge Long Ju, Zhengguan Lu, Tonghang Han, Jixiang Yang, T. Senthil, Trithep Devakul, Yongxin Zeng, Patrick J. Ledwith, Eslam Khalaf, Ruihua Fan, Rahul Sahay, Erez Berg, Dmitry Chichinadze, Chang-Geun Oh, Charlie Marcus, Maksym Serbyn, and Bertrand I. Halperin for insightful discussions.
T.W., T.W., and M.Z. are supported by the U.S. Department of Energy, Office of Science, Office of Basic Energy Sciences, Materials Sciences and Engineering Division under Contract No. DE-AC02-05-CH11231 (Theory of Materials program KC2301). T.W. is also supported by the Heising-Simons Foundation, the Simons Foundation, and NSF grant No. PHY-2309135 to the Kavli Institute for Theoretical Physics (KITP). A.V. is supported by the Simons Collaboration on Ultra-Quantum Matter, which is a grant from the Simons
Foundation (651440, A.V.) and by the Center for Advancement of Topological Semimetals, an Energy Frontier Research Center funded by the US Department of
Energy Office of Science, Office of Basic Energy Sciences,
through the Ames Laboratory under contract No. DEAC02-07CH11358. 
This research is funded in part by the
Gordon and Betty Moore Foundation’s EPiQS Initiative,
Grant GBMF8683 to T.S.  D.E.P. is supported by the Simons Collaboration on Ultra-Quantum Matter, which is a grant from the Simons Foundation.
\end{acknowledgements}

\bibliographystyle{unsrt}
\bibliography{references,CFL_bib,footnotes,taige_bib}
\onecolumngrid
\appendix

\section{Derivation of Hartree and Fock energies for RMG spinors}
\label{app:RMG_Hartree_Fock}
We evaluate the Hartree and Fock energies Eqs.~(\ref{eq:RMG_Hartree},\ref{eq:RMG_Fock}) explicitly. Recall that the spinor is given by

\begin{equation}
    \chi^{(N_L)}_{n+1} \coloneqq\chi^{(N_L)}(\vv{\kappa}_{n+1}) = \mathcal{N}(1, \eta e^{2\pi in/6}, \eta^2 e^{4\pi in/6},\ldots, \eta^{N_L-1}e^{2\pi i(N_L-1)n/6}),
\end{equation}
where $\mathcal{N} = 1/\sqrt{\sum_{a=0}^{N_L - 1} \eta^{2a}}$.

The form factor is defined by $\Lambda_{n, m} = \chi^\dagger_n \chi_m$. Then,

\begin{align}
    \Lambda_{3, 1} &= \Lambda_{6, 4} = \mathcal{N}^2 \sum_{n=0}^{N_L -1 } \eta^{2n} e^{-2\pi in/3},\\
    \Lambda_{1, 2} &= \Lambda_{4, 5} = \mathcal{N}^2 \sum_{n=0}^{N_L -1 } \eta^{2n} e^{2\pi in/6}.
\end{align}

Therefore, we get

\begin{align}
    \epsilon_H(C) &= \frac{2}{3} \mathrm{Re}[e^{2\pi iC/3} \Lambda_{1,3} \Lambda_{4,6}] \\
    &=\frac{2}{3} \mathrm{Re}\left[e^{2\pi iC/3} \left(\mathcal{N}^2 \sum_{n=0}^{N_L -1 } \eta^{2n} e^{2\pi in/3}\right)^2\right]\\
    &=\frac{2}{3} \mathcal{N}^4 \sum_{a, b=0}^{N_L-1} \eta^{2(a+b)} \mathrm{Re}[e^{2\pi iC/3}   e^{i2(a+b)\phi}] \\
    &= \frac{2}{3} \mathcal{N}^4 \sum_{a, b=0}^{N_L-1} \eta^{2(a+b)} \cos\left(\frac{2\pi (a+b+C)}{3}\right)
\end{align}

Similarly, we get for Fock

\begin{align}
    \epsilon_F(C) &= -\frac{2}{3} \mathrm{Re}[e^{-2\pi iC/3} \Lambda_{1, 2} \Lambda_{4, 5}] \\
    &=-\frac{2}{3} \mathrm{Re}\left[e^{-2\pi iC/3} \left(\mathcal{N}^2 \sum_{n=0}^{N_L -1 } \eta^{2n} e^{2\pi in/3}\right)^2\right]\\
    &=-\frac{2}{3} \mathcal{N}^4 \sum_{a, b=0}^{N_L-1} \eta^{2(a+b)} \mathrm{Re}[e^{-2\pi iC/3}   e^{2\pi i(a+b)/3}] \\
    &= -\frac{2}{3} \mathcal{N}^4 \sum_{a, b=0}^{N_L-1} \eta^{2(a+b)} \cos\left(\frac{2\pi(a+b-2C)}{6}\right)
\end{align}

\section{Proof of degeneracy under contact interaction}
\label{app:two_component_degenerate}
We will now prove the curious fact that if $\kappa$ and $\kappa'$ spinors only have two components (with angular momentum $0$ and $1$, respectively), then the Chern number $0$ and $1$ states will remain degenerate under contact interaction.

We now define
\begin{equation}
    \chi(\vv\kappa_1) = \begin{pmatrix}
        a_{\kappa}\\
        b_{\kappa}
    \end{pmatrix},\quad
    \chi(\vv\kappa_4) = \begin{pmatrix}
        a_{\kappa'}\\
        b_{\kappa'}
    \end{pmatrix}.
\end{equation}
The $C_3$ action is given again by $\hat{C}_3 = \mathrm{diag}(1,e^{2\pi i/3})$. We fix the gauge such that $\chi(\vv \kappa_{j+2})=\hat{C}_3 \chi(\vv \kappa_j)$.

Making use of Eqs.~(\ref{eq:Hartree_FF},\ref{eq:Fock_FF}), we obtain
\begin{equation}
\begin{aligned}
    \epsilon_H(C) &= \frac{2}{3} \mathrm{Re}\left[e^{2\pi iC/3}\Lambda_{1,3}\Lambda_{4,6}\right]\\
    &=\frac{2}{3} \mathrm{Re}\left[e^{2\pi iC/3}(|a_{\kappa}|^2+e^{2\pi i/3}|b_{\kappa}|^2)(|a_{\kappa'}|^2+e^{2\pi i/3}|b_{\kappa'}|^2)\right]\\
    &=\frac{2}{3}\left(|a_{\kappa}|^2|a_{\kappa'}|^2 \cos(2\pi C/3)+(|a_{\kappa}|^2|b_{\kappa'}|^2+|b_{\kappa}|^2|a_{\kappa'}|^2)\cos(2\pi (C+1)/3)+|b_{\kappa}|^2|b_{\kappa'}|^2\cos(2\pi (C+2)/3)\right),
\end{aligned}
\end{equation}
\begin{equation}
\begin{aligned}
    \epsilon_F(C) &= -\frac{2}{3} \mathrm{Re}\left[e^{-2\pi iC/3}\Lambda_{1,2}\Lambda_{4,5}\right]\\
    &=-\frac{2}{3} \mathrm{Re}\left[e^{-2\pi iC/3}(\overline{a_{\kappa}} a_{\kappa'}+e^{-2\pi i/3}\overline{b_{\kappa}} b_{\kappa'})(\overline{a_{\kappa'}} a_{\kappa}+e^{-2\pi i/3}\overline{b_{\kappa'}} b_{\kappa})\right]\\
    &=-\frac{2}{3}\left(|a_{\kappa}|^2|a_{\kappa'}|^2 \cos(2\pi C/3)+(\overline{a_{\kappa'}} a_{\kappa}\overline{b_{\kappa}} b_{\kappa'}+\overline{a_{\kappa}} a_{\kappa'}\overline{b_{\kappa'}} b_{\kappa})\cos(2\pi (C+1)/3)+|b_{\kappa}|^2|b_{\kappa'}|^2\cos(2\pi (C+2)/3)\right)
\end{aligned}
\end{equation}

In contact interactions, we can take $V_{\vv g}=V_{\kappa}=1$. Thus
\begin{equation}
\begin{aligned}
    E(C) &= \epsilon_H(C)+\epsilon_F(C)= \frac{2}{3}\cos\left(\frac{2\pi(C+1)}{3}\right)|\overline{a_{\kappa}} a_{\kappa'}-\overline{b_{\kappa}} b_{\kappa'}|^2
\end{aligned}
\end{equation}
which means that $E(0)=E(1)$. We note that contact interaction annihilates all intra-orbital interactions, and one can generally prove that the $C=0$ and $C=1$ states are still degenerate under any inter-orbital interaction.

\section{Real space properties of orbitals at $\kappa$ and $\kappa'$}
\label{app:real_space_properties}

\begin{figure}
    \centering
    \includegraphics[width=0.49\textwidth]{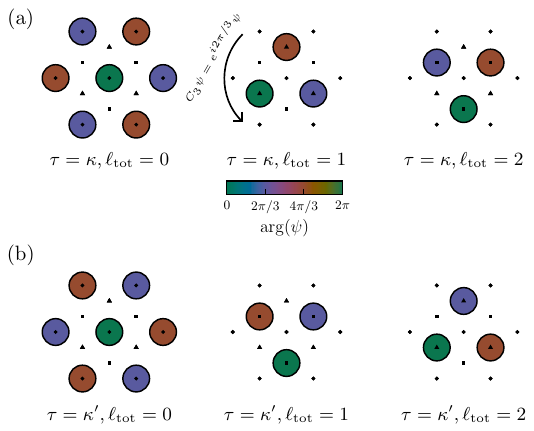}
    \caption{Real-space plot of high-symmetry point wavefunctions $\phi_{\tau,\ell_{\mathrm{tot}}}(\vv r)$. The size of the circles corresponds to the amplitudes of the wavefunctions at the Wyckoff points, and the color corresponds to the phase of the wavefunctions. The rows correspond to different high symmetry points (a) $\kappa$ and (b) $\kappa'$, and the columns correspond to different angular momenta $\ell_{\mathrm{tot}}$. Different wavefunctions at the same high-symmetry point are related by translations.}
    \label{fig:high-symmetry-wavefunction}
\end{figure}

In this appendix, we will explain two elementary characteristics of $C_3$ symmetric orbitals at $\kappa$ and $\kappa'$ points. Recall that Wyckoff positions are points in the unit cell that are invariant under a spatial symmetry action. With the current choice of the lattice, these points are integer multiples of
\begin{equation}
    \tilde{\vv a} = \frac{1}{3} (\vv a_1+\vv a_2).
\end{equation}

Below we will establish the following two facts:

\begin{enumerate}
    \item The orbital-resolved charge densities of $C_3$ symmetric states at $ \kappa$ and $ \kappa'$ points have peaks and zeros at the Wyckoff points, determined by both the orbital angular momentum $\ell_a$ and the state angular momentum $\ell_{\xi}$;
    \item The action of the translation $T_{\tilde{\vv a}}$ translates $\ell_{\kappa}\to \ell_{\kappa}+1$, and $\ell_{\kappa'}\to \ell_{\kappa'}-1$.
\end{enumerate}

To understand the first statement, take the $\vv\kappa$ point without loss of generality. Now, consider the state $\psi_{\kappa,\ell_{\kappa}}(\vv r)$: its orbital-resolved charge density is given by

\begin{equation}
    \rho_{\kappa}^{a}(\vv{r}, \ell_{\kappa})
=\n{\psi^a_{\kappa,\ell_{\kappa}}(\vv{r})}^2.
\end{equation}

We now make use of the definition Eq.~\eqref{eq:three_patch_wavefunctions}. Consider the momentum point $\vv\kappa_1$: Let's say $\chi^a(\vv\kappa_1) = \chi_0$. Then, due to the gauge choice $\chi(\vv\kappa_3) = \hat{C}_3 \chi(\vv\kappa_1)$, we have $ \chi^a(\vv\kappa_3)=\exp[2\pi i\ell_a/3]\chi_0$, and similarly $\chi^a(\vv\kappa_5)=\exp[4\pi i\ell_a/3]\chi_0$.

We may now compute the angular-momentum resolved charge density:

\begin{equation}
    \begin{aligned}
        \rho_{\kappa}^{\ell_a}(\vv{r}, \ell_{\kappa})
    &=\n{\psi^a_{\kappa,\ell_{\kappa}}(\vv{r})}^2\\
    &=\frac{1}{3}\n{\sum_{j=1,3,5}\exp{\left[\frac{2\pi i \ell_{\kappa}}{3}\frac{j}{2}\right]}e^{i\vv \kappa_j\cdot \vv r}\chi(\vv \kappa_j)^a}^2\\
    &=\frac{\n{\chi(\vv \kappa_1)^a}^2}{3}\n{\sum_{j=1,3,5}\exp{\left[\frac{2\pi i (\ell_a+\ell_{\kappa})}{3}\frac{j}{2}\right]}e^{i\vv \kappa_j\cdot \vv r}}\\
    &=\n{\chi(\vv \kappa_1)^a}^2 \n{\varphi_{\kappa,\ell_{\kappa}+\ell_{a}}(\vv r)}^2
    \end{aligned}
\end{equation}

Since $\vv\kappa_j\cdot \tilde{\vv a}= 2\pi j/3$ for $j\in 1,3,5$ and $-2\pi j/3$ for $j=2,4,6$ (modulo $2\pi$), we may immediately deduce

\begin{equation}
    \rho_{\kappa}^{a}(m\tilde{\vv a}, \ell_{\kappa}) = 3\n{\chi(\vv \kappa_1)^a}^2 \delta_{m, \ell_a+\ell_{\kappa}}.
\end{equation}
where the $\delta_{m,n}=1$ if $3|(m-n)$ and $0$ otherwise.

This motivates a definition of 
\begin{equation}
    \ell_{\mathrm{tot}} = \ell_a+\ell_{\kappa}.
\end{equation}
Indeed, a similar calculation for the $\kappa'$ shows that

\begin{equation}
    \rho_{\tau}^{a}(m\tilde{\vv a}, \ell_{\tau}) \propto \delta_{m, \tau\ell_{\mathrm{tot}}}.
\end{equation}
where $\tau=\pm$ labels the $\kappa$ and $\kappa'$ points.

This establishes the first fact: the peak of the charge density is at a Wyckoff position determined by a signed total angular momentum, $\tau(\ell_a+\ell_\tau)$. Particularly, one can see in Fig.~\ref{fig:high-symmetry-wavefunction} that although the charge density peaks of $\ell_{\mathrm{tot}}=1$ at $\kappa$ point and $\ell_{\mathrm{tot}}=-1$ at $\kappa'$ point share the same Wyckoff positions, the wavefunctions have different phase structures and thus different angular momenta.

As for the second fact, we show it explicitly:
\begin{equation}
    \begin{aligned}
    (T_{\tilde{\vv a}}\psi_{\kappa})(\vv r, \ell_{\kappa}) &= \psi_{\kappa}(\vv r-\tilde{\vv a}, \ell_{\kappa})\\
    &= \frac{1}{\sqrt{3}}\sum_{j=1,3,5} 
    \exp{\left[\frac{2\pi i \ell_{\kappa}}{3}\frac{j}{2}\right]}
    e^{i\vv \kappa_j\cdot (\vv r-\tilde{\vv a})}\chi(\vv \kappa_j) \\
    &\propto \frac{1}{\sqrt{3}}\sum_{j=1,3,5} 
    \exp{\left[\frac{2\pi i (\ell_{\kappa}+1)}{3}\frac{j}{2}\right]}
    e^{i\vv \kappa_j\cdot \vv r}\chi(\vv \kappa_j)\\
    &=\psi_{\kappa}(\vv r, \ell_{\kappa}+1).
    \end{aligned}
\end{equation}
The $\kappa'$ point calculation is exactly similar. Both of these statements can be checked graphically in Fig.~\ref{fig:high-symmetry-wavefunction}.

\section{Justification of the three-patch model}
\label{sec:justifications}

In this appendix, we will motivate the building blocks of the three-patch model from simple physical conditions of the single-particle band structure. We will also discuss how we arrive at the restricted Hilbert space in Eq.~\eqref{eq:three_patch_wavefunctions} and below.

Spontaneous translation symmetry breaking can occur in the case that the interaction strength $U$ is much larger than the bandwidth of the lowest band within the mBZ:
\begin{equation}
    |E({\vv\gamma}) - E({\vv\kappa_i})| \lesssim U
\end{equation}
In the limit where the lowest band is completely flat, the Fermi surface can be arbitrarily deformed without being penalized by the kinetic energy. In this way, we can create low-energy Fermi surfaces with perfect nesting conditions, which are in general unstable to the creation of insulating states by the exchange interaction. This thus corresponds to condition (i) stated at the beginning of Sec.~\ref{sec:three-patch-model}. Once we take the dispersion into account, a gap equation must be solved to investigate the existence of such instabilities, which is performed under various approximations in Sec.~\ref{sec:gap_equation}.

The ``weak coupling mean field'' assumption has two subparts. It firstly states that the state is weakly coupled. This can also be implemented by an energetic constraint: we need a large gap between the $\vv{\gamma}$ point and its higher harmonics, which means that there will be a large direct gap around the $\vv\gamma$ point in the folded band structure. In the case where
\begin{equation}
    E(\vv{\gamma} + \vv{g}_i) - E(\vv{\gamma}) > U,
\end{equation}
interactions cannot hybridize the $\vv\gamma$ point wavefunction with other plane waves, and thus the state will be weakly coupled. It also states that the state can be described as a Slater determinant state. While this mean-field approximation is uncontrolled, as long as we consider a variational space of Hartree-Fock states, this assumption is always satisfied.

The three-patch assumption is harder to justify \textit{a priori} from the knowledge of the single-particle Hamiltonian $\hat{h}$ alone. Physically, the Fock interaction is strong between nearby momentum given the small momentum exchange, and thus the single particle wavefunctions can be plausibly inferred to be similar. However, realistically, the best we can do is an \textit{a posteriori} justification by inspecting the nature of the SCHF ground state, which we have performed carefully for the ground states of the RMG Hamiltonian.

In the remainder of this section, we will explain in detail how to physically obtain the reduced Hilbert space Eq.~\eqref{eq:three_patch_wavefunctions} from the assumptions above.

We only consider Slater determinant states. After we fold the Brillouin zone, the states are then labeled by crystalline momentum $\vv k$:
\begin{equation}
    \ket{\Psi} =\prod_{\vv{k}} \prod_{j=1}^{n_{\vv k}} d^\dagger_{\vv{k}, j} \ket{0},
\end{equation}
where $n_{\vv{k}}$ is the occupation number at mBZ momentum $\vv{k}$, and $d^\dagger_{\vv{k}, j}$ is the creation operator with mBZ momentum $\vv{k}$. Since we assumed the state to be an insulator, we fill the low-energy band of the mean-field Hamiltonian in the mBZ such that $n_{\vv k}=1$ for all crystalline momentum $\vv{k}$.

The weak-coupling assumption then allows us to reduce to a low-energy Hilbert space with the lowest energy plane waves. 
Explicitly, we can restrict to a Slater determinant states generated by low energy plane waves:
\begin{equation}
    \label{eq:folded_BZ_operators}
    \hat{d}^\dagger_{\k n} = \sum_{\g} \psi^{\g}_{\k n} \hat{c}^\dagger_{\k + \g}.
\end{equation}
Here the sum over $\g$ is only non-zero when $E(\k+\g) - E(\k)$ is small relative to the interaction strength.

Finally, the three-patch enables a further reduction to an analytically tractable model. Since the wavefunction is assumed to be homogeneous within each patch, we pick one representative momentum to describe the whole patch. The most natural choice is the set of high symmetry points $\xi = \st{{\gamma}, {\kappa}, {\kappa}'}$. Schematically, this reduction in Hilbert space can be written as:

\begin{equation}
    \ket{\Psi} = \prod_{\vv k\in \mathrm{mBZ}}d^\dagger_{\vv k}\ket{0}\xrightarrow[\text{transformation}]{\text{three-patch}}\prod_{\substack{\vv k \in \\ \{\gamma, \kappa, \kappa'\}}}d^\dagger_{\vv k}\ket{0}=\ket{\Psi_{\mathrm{3p}}}
\end{equation}

The high symmetry points $\st{{\gamma}, {\kappa}, {\kappa}'}$ have one, three, and three low-energy states respectively. We will assume that electronic states from the higher shells have a gap to these low-energy states that is much larger compared to the interaction energy.
Following \eqref{eq:folded_BZ_operators}, and $C_3$ symmetry, the associated creation operators are
$\hat{d}^\dagger_{{\gamma}}, \hat{d}^\dagger_{{\kappa}, \ell_{{\kappa}}}$, and $ \hat{d}^\dagger_{{\kappa}', \ell_{{\kappa}'}}$ with $C_3$ angular momentum $\ell_{\xi}$. Their single-particle wavefunctions $\psi(\r) = \braket{\r|\hat{d}^\dagger|0}$ are thus given by Eq.~\eqref{eq:three_patch_wavefunctions}, which we reproduce here
\begin{align}
\label{eq:three_patch_wavefunctions_simpler}
    \psi_{ \gamma,\ell_{\gamma}=0} &= \chi(\gamma) \\
    \nonumber 
    \psi_{ \kappa,\ell_{ \kappa}}(\vv r) &= \frac{1}{\sqrt{3}}\sum_{j=1,3,5} 
    \exp{\left[\frac{2\pi i \ell_{\kappa}}{3}\frac{j}{2}\right]}
    e^{i\vv \kappa_j\cdot \vv r}\chi(\vv \kappa_j),\\
    \nonumber
    \psi_{ \kappa',\ell_{ \kappa'}}(\vv r) &= \frac{1}{\sqrt{3}}\sum_{j=2,4,6} 
    \exp{\left[\frac{2\pi i \ell_{\kappa'}}{3}\frac{j}{2}\right]}
    e^{i\vv \kappa_j\cdot \vv r}\chi(\vv \kappa_j).
\end{align}
By the assumption that we have an insulating state with $\nu=1$, we have one electron per representative state in the patch. Together with the Slater determinant assumption, we arrive at variational states

\begin{equation}
    \Psi_{\ell_{{\kappa}}, \ell_{{\kappa}'}}(\vv{r}_1, \vv{r}_2, \vv{r}_3) = \mathcal{A}[\psi_{ \gamma}(\vv{r}_1), \psi_{ \kappa,\ell_{ \kappa}}(\vv{r}_2),  \psi_{ \kappa',\ell_{\vv \kappa'}}(\vv{r}_3)].
\end{equation}

This establishes the three-patch model.

\section{Energetic competition of the three-patch states}
\label{app:three-patch-energy-competition}

\begin{figure}
    \centering
        \includegraphics[width=0.7\linewidth]{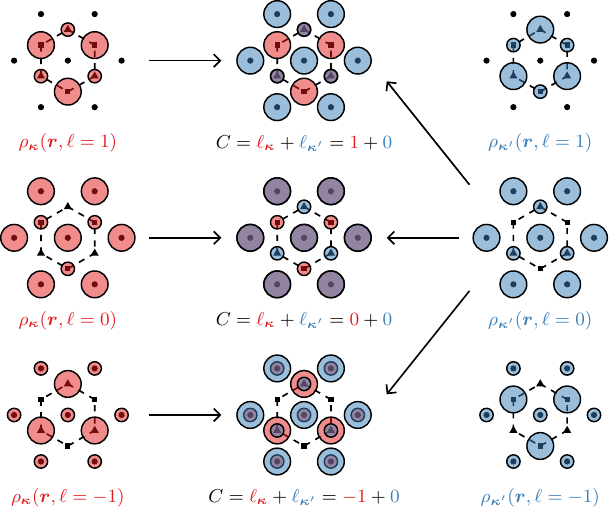}
    \caption{Reproduction of Fig.~\ref{fig:3point-model-hartree}. Charge densities of competing candidate states of the three-patch model. (left): the charge densities at the $\kappa$ point. The size of the circles corresponds to different amplitudes of charge densities, which come from different components of the spinor. Different angular momentum states labeled by $\ell$ are related to each other by translation. (right): the charge densities in the ${\kappa}'$ point. (center): the total charge densities of states with different Chern numbers.}
    \label{fig:3point-model-hartree_app}
\end{figure}

We now use the charge density pattern above to infer some features of the Hartree and Fock energies. The Hartree energy between $\psi_{\kappa}$ and $\psi_{\kappa'}$ corresponds to the interaction energy of classical charge configurations. It now becomes a simple visual exercise: we can visually figure out which charge densities are on top of each other by looking at Fig.~\ref{fig:3point-model-hartree_app}. It is easy to see that the $C=1$ state has the least density overlap between $\rho_{{\kappa}}, \rho_{{\kappa}'}$. This is confirmed in the Hartree energy plot in Fig.~\ref{fig:3point-model-hartree_app}.

At small $\eta$, the Fock term can be similarly graphically evaluated. This is because Fock exchange energy is controlled by the alignment of pseudospins, and at small $\eta$ most weight is on the top layer. We conclude the $C=0$ state has the lowest Fock energy due to the large overlap between the top layer charge densities, as confirmed in the Fock energy plot in Fig.~\ref{fig:3point-model-hartree_app}. We note, however, that this visual exercise misses other contributions such as intercomponent interactions, and is qualitatively incorrect at large $\eta$, as evidenced by the fact that $C=1$ state can have lower Fock energy.

We note an interesting implication of the density patterns. Consider a  potential such that $V(\vv r) = V_0\sum_{i} \cos(\vv g_i\cdot \vv r)$, where $V_0<0$. This potential has a honeycomb shape, where the electrons are favored to form a triangular lattice at the superlattice points. If we apply this potential to the first component of the spinor, the $C=0$ state, which has a triangular charge density in the first component, gains more energy than $C=\pm 1$ states.
On the other hand, if the potential is applied to the second component, the $C=1$ state is favored over the other states, since the charge density of the second component forms a triangular lattice.
This consideration implies that the phase boundary can be shifted by such external potentials.

\section{The three-patch model using realistic spinors}
\label{app:realistic_spinors}

\begin{figure}
    \centering
    \includegraphics[width=\textwidth]{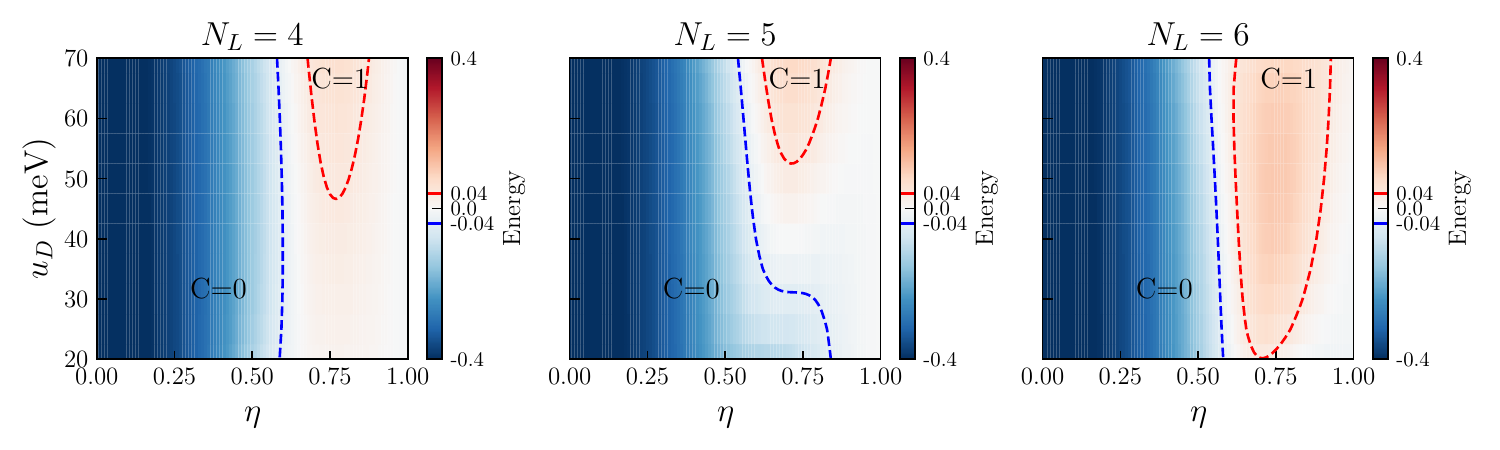}
    \caption{The energy difference between $C=0$ and $C=1$ states using the realistic spinors with the three-patch model, as described in the App.~\ref{app:realistic_spinors}. The blue and red dashed lines correspond to the contour lines with energy $-0.1$ and $0.1$ respectively. }
    \label{fig_app:realistic spinors}
\end{figure}

In the main text, we used the sublattice projected spinor $\chi_B$ (Eq.~\eqref{eq:spinor}) to compute the energetic competition between different candidate states. In this section, we consider the three-patch model with the full spinor generated by the single particle Hamiltonian Eq.~\eqref{eq:RMG_fullham}, which takes the presence of the other sublattice $A$ into account as well. We note a similar computation was done in Ref.~\onlinecite{dong_stability_2024}.

In Fig.~\ref{fig_app:realistic spinors}, we show the Coulomb energy difference between the $C=1$ and $C=0$ states, computed using the three-patch model. The input spinors for the three-patch model were taken to be the eigenstates of the single particle Hamiltonian of RMG at different values of $N_L$. As in Fig.~\ref{fig:schematic_phase_diagram}, the spinors are taken to be at the corners of the BZ, whose size is parametrized by $\eta = v_F|\vv{\kappa}_i(n) - \vv{K}|/t_1$. 

As shown in Fig.~\ref{fig:schematic_phase_diagram}, smaller $\eta$ stabilizes the $C=0$ state consistently for different values of $N_L$. At higher $\eta$, the $C=1$ state becomes more favorable, consistent with the computation using $\chi_B$. In particular, at large $u_D$, the boundary between $C=0$ and $C=1$ is in quantitative agreement with that computed from $\chi_B$, since the large displacement field suppresses mixing between the two zero modes. On the other hand, at small $u_D$, there is a significant deviation from the prediction from $\chi_B$, especially at $N_L=5$.

We note, however, that the small $u_D$ region is part of a large ``gray area'' where the energy difference between the two competing states is small. In this region, we should not take the phase diagram from the three-patch model at face value, as second-order effects neglected in our modelling likely determine the ground state in this regime.

\section{Dependence of the phase diagram on gate distance}
\label{app:gate_dependence}

In this Appendix, we examine the effect of gate distance on the phase diagram of hombohedral pentalayer graphene.
This allows us to trace the evolution of the phase diagram from the Coulomb limit to the contact limit as the gate distance decreases discussed in Fig.~\ref{fig:three-patch_hartree-fock_comparison} of the main text.

Recall that the double-gated Coulomb interaction is given by $V_{\vv{q}}=2\pi\tanh|\vv{q}|d/(\epsilon_r \epsilon_0 |\vv{q}|)$, where $d$ is the distance to the metallic gates. In the long-distance limit, $V_{\vv{q} \to 0} = 2\pi d / (\epsilon_r \epsilon_0)$. In the following calculation, we use fixed $d/\epsilon_r$ so that $V_{\vv{q}\to 0}$ remains constant.

\begin{figure}
    \centering
    \includegraphics[width=\linewidth]{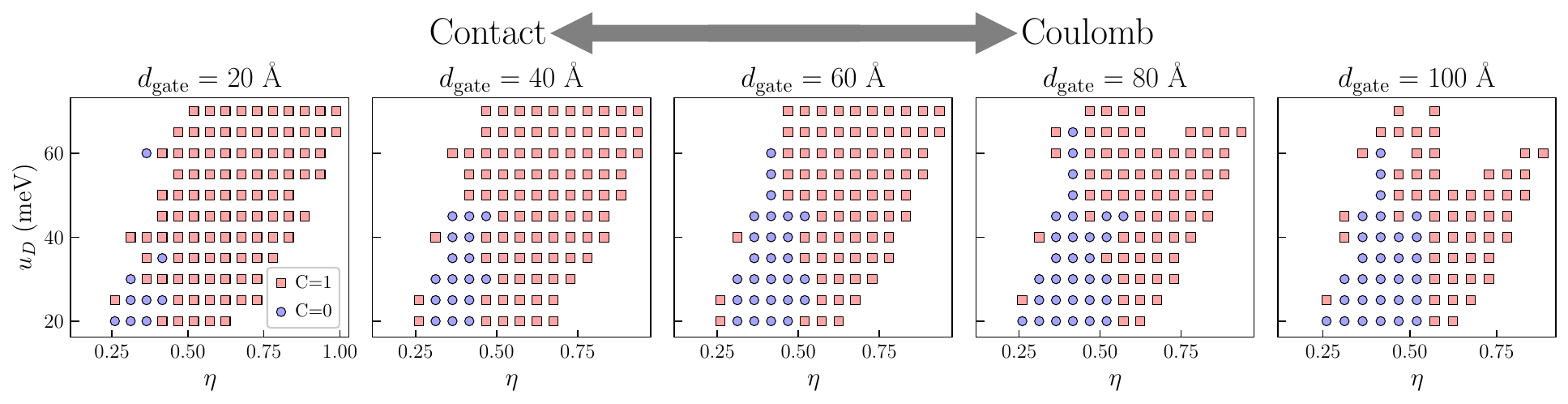}
    \caption{The phase diagram of rhombohedral pentalayer graphene at various gate distances, computed using SCHF. The interaction strength used is $V_{\vv{q} \to 0} =73.55 \si{eV}\cdot\si{nm}$, or $\epsilon_r = d/d_0$ where $d_0 = 2 \si{nm}$. Missing datapoints correspond to metallic states.
    }
    \label{fig_app:gate_dependence}
\end{figure}

In Fig.~\ref{fig_app:gate_dependence}, we show the phase diagram of rhombohedral pentalayer graphene at various gate distances. Towards the contact limit (small gate distances), the $C=1$ state relative to the $C=0$ state --- consistent with the prediction of the three-patch model.

\begin{figure}
    \centering
    \includegraphics[width=\linewidth]{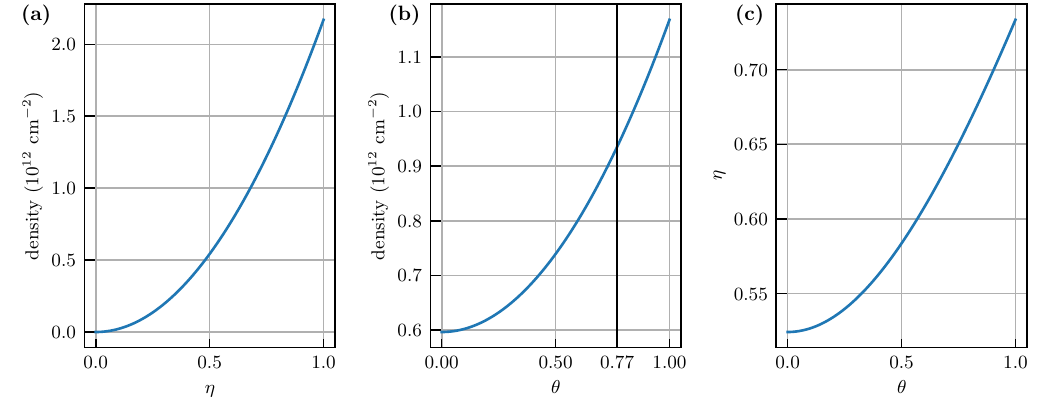}
    \caption{Conversions between the dimensionless parameter $\eta$ and experimentally measurable quantities. (a) Electron density versus $\eta$. (b) Electron density versus the moir\'e twist angle $\theta$. (c) $\eta$ versus $\theta$.}
    \label{fig:eta_theta_conversion}
\end{figure}

To facilitate comparison between these results and experiments, Fig. \ref{fig:eta_theta_conversion} shows the correspondence between the dimensionless parameters $\eta = \frac{v_F \n{\vv{\kappa}_i - \vv{K}}}{t_1}$ and experimental parameters of charge density and twist angle. For instance, $\theta=0.77^\circ$ corresponds to a charge density $n \approx \SI{0.93e12}{\per\centi\meter\squared}$, which in turns corresponds to $\eta \approx 0.66$.

\end{document}